\documentclass[showpacs,preprintnumbers,amsmath,amssymb,floatfix]{revtex4}
\usepackage{epsf}
 \newcommand{\insertplot}[5]{\begin{figure}
 \hfill\hbox to 0.05in{\vbox to #5in{\vfill
 \inputplot{#1}{#4}{#5}}\hfill}
 \hfill\vspace{-.1in}
 \caption{#2}\label{#3}
 \end{figure}}
 \newcommand{\inputplot}[3]{
 \special{ps: plotfile #1}
}

\usepackage[german, english]{babel}
\usepackage{ifthen}
\usepackage{epsfig}
\newcounter{fig}   \newcommand{\lbfig}[1]{\refstepcounter{fig}
\label{#1} }

\newcommand{\vphi}{\varphi}

\begin{document}
\title{Rotating Electroweak Sphaleron-Antisphaleron Systems}
\author{
{\bf Rustam Ibadov}
}
\affiliation{Department of Theoretical Physics and Computer Science,\\
Samarkand State University, Samarkand, Uzbekistan}
\author{
{\bf Burkhard Kleihaus, Jutta Kunz and Michael Leissner}
}
\affiliation{
{Institut f\"ur Physik, Universit\"at Oldenburg, 
D-26111 Oldenburg, Germany}
}
\date{\today}

\begin{abstract}
At finite weak mixing angle
the sphaleron solution of Weinberg-Salam theory
can be endowed with angular momentum
proportional to the electric charge.
Here we show, that this holds also 
for sphaleron-antisphaleron systems
such as pairs, chains and vortex rings.
We also address the equilibrium conditions
for these solutions.
\end{abstract}

\maketitle

\section{Introduction}

The electroweak sector of the standard model 
harbours unstable non-perturbative classical solutions: sphalerons
\cite{Manton:1983nd,Klinkhamer:1984di,Kleihaus:1991ks,Kunz:1992uh}.
The static sphaleron solution of Weinberg-Salam theory represents
the top of the energy barrier between topologically inequivalent vacua.
Since the standard model does not absolutely conserve
baryon number \cite{'tHooft:1976up,Ringwald:1989ee}, at finite temperature
baryon number violating processes can arise because of
thermal fluctuations of the fields
large enough to overcome the energy barrier between 
distinct vacua.
The rate for baryon number violating processes
is then largely determined by a Boltzmann factor,
containing the height of the barrier at a given temperature
\cite{McLerran:1993rv,Rubakov:1996vz,Klinkhamer:2003hz,Dine:2003ax}.
The sphaleron itself carries baryon number $Q_{\rm B}=1/2$ 
\cite{Klinkhamer:1984di}.

At finite weak mixing angle the static electroweak sphaleron
possesses a large magnetic moment but does not carry electric charge.
As argued before \cite{Saffin:1997ae}
and demonstrated recently in nonperturbative studies
\cite{Radu:2008ta,Kleihaus:2008cv},
the addition of electric charge 
leads
to a non-vanishing Poynting vector and thus a finite angular momentum
density of the system. 
Consequently 
a branch of spinning electrically charged sphalerons arises.
Since these carry non-vanishing baryon number as well,
they also entail baryon number violating processes.

Beside the sphaleron,
the non-trivial topology of the configuration space of Weinberg-Salam theory
gives rise to further unstable classical solutions.
A superposition of $n$ sphalerons, for instance,
can lead to static axially symmetric solutions, multisphalerons, 
which carry baryon number $Q_{\rm B}=n/2$ and
whose energy density is torus-like 
\cite{Brihaye:1994ib,Kleihaus:1994yj,Kleihaus:1994tr}.
A superposition of a sphaleron and an antisphaleron,
on the other hand, can give rise to a bound
sphaleron-antisphaleron system, in which a sphaleron and an
antisphaleron are located at an equilibrium distance
on the symmetry axis 
\cite{Klinkhamer:1985ki,Klinkhamer:1990ik,Klinkhamer:1993hb}.
Such a sphaleron-antisphaleron pair has vanishing baryon number,
$Q_{\rm B}=0$, since the antisphaleron carries $Q_{\rm B}=-1/2$.
The sphaleron-antisphaleron pair therefore does not mediate
baryon number violating processes.

Recently, 
the sphaleron-antisphaleron pair solutions have been generalized,
to form sphaleron-antisphaleron chains,
where $m$ sphalerons and antisphalerons are located
on the symmetry axis, in static equilibrium
\cite{Kleihaus:2008gn},
in close analogy to the monopole-antimonopole chains
encountered in the Georgi-Glashow model 
\cite{Kleihaus:2003nj}.
When multisphalerons and -antisphalerons
are considered instead, vortex ring solutions arise,
when $n \ge 3$ \cite{Kleihaus:2008gn},
where the Higgs field vanishes not (only) on isolated points
on the symmetry axis but (also) on one or more rings,
centered around the symmetry axis 
\cite{Kleihaus:2003xz,Kleihaus:2004is}.

Here we consider 
the addition of electric charge 
to such multisphaleron and sphaleron-antisphaleron systems.
As for the simple sphaleron,
the non-vanishing Poynting vector leads to a finite angular momentum
density for these configurations.
Thus branches of rotating electrically charged sphalerons
emerge from the respective static electrically neutral
configurations.
We construct these solutions explicitly and discuss their properties.
The angular momentum and the electric charge of the solutions
are proportional \cite{VanderBij:2001nm,Radu:2008ta,Kleihaus:2008gn}.

In section 2 we present the action, the Ansatz and the
boundary conditions. 
In section 3 we consider the
relevant physical properties and, in particular, 
derive the linear relation
between angular momentum and electric charge.
We present and discuss the numerical results in section 4.
These include global properties of the solutions,
such as their energy, their angular momentum, their charge
and their magnetic moments,
but also local properties, such as their
energy density, their angular momentum density
and the modulus of their Higgs field.
Moreover, we discuss the equilibrium conditions
for the solutions. 
We give our conclusions in section 5.

\section{Action and Ansatz}


We consider the bosonic sector of Weinberg-Salam theory
\begin{equation}
{\cal L} = -\frac{1}{2} {\rm Tr} (F_{\mu\nu} F^{\mu\nu})
-  \frac{1}{4}f_{\mu \nu} f^{\mu \nu}                                           
- (D_\mu \Phi)^{\dagger} (D^\mu \Phi) 
- \lambda (\Phi^{\dagger} \Phi - \frac{v^2}{2} )^2 
\  
\label{lag1}
\end{equation}
with su(2) field strength tensor
\begin{equation}
F_{\mu\nu}=\partial_\mu V_\nu-\partial_\nu V_\mu
            + i g [V_\mu , V_\nu ]
 , \end{equation}
su(2) gauge potential $V_\mu = V_\mu^a \tau_a/2$,
u(1) field strength tensor
\begin{equation}
f_{\mu\nu}=\partial_\mu A_\nu-\partial_\nu A_\mu 
 , \end{equation}
and covariant derivative of the Higgs field
\begin{equation}
D_{\mu} \Phi = \Bigl(\partial_{\mu}
             +i g  V_{\mu} 
             +i \frac{g'}{2} A_{\mu} \Bigr)\Phi
 , \end{equation}
where $g$ and $g'$ denote the $SU(2)$ and $U(1)$ gauge coupling constants,
respectively,
$\lambda$ denotes the strength of the Higgs self-interaction and
$v$ the vacuum expectation value of the Higgs field.

The gauge symmetry is spontaneously broken 
due to the non-vanishing vacuum expectation
value of the Higgs field
\begin{equation}
    \langle \Phi \rangle = \frac{v}{\sqrt2}
    \left( \begin{array}{c} 0\\1  \end{array} \right)   
 , \label{Higgs} \end{equation}
leading to the boson masses
\begin{equation}
    M_W = \frac{1}{2} g v   , \ \ \ \ 
    M_Z = \frac{1}{2} \sqrt{(g^2+g'^2)} v  \, , \ \ \ \ 
    M_H = v \sqrt{2 \lambda} \,  . 
\end{equation}
$ \tan \theta_{\rm w} = g'/g $ determines
the weak mixing angle $\theta_{\rm w}$,
defining the electric charge $e = g \sin \theta_{\rm w}$.  
We also denote the weak fine structure constant $\alpha_{\rm W}=g^2/4\pi$.


To obtain stationary rotating solutions of the bosonic sector
of Weinberg-Salam theory,
we employ the time-independent axially symmetric Ansatz
\begin{equation}
V_\mu\, dx^\mu
  = \left( B_1\, \frac{\tau^{(n,m)}_r}{2g} 
         + B_2\, \frac{\tau^{(n,m)}_\theta}{2g} \right) \, dt 
            -n\sin\theta\left(H_3 \frac{\tau^{(n,m)}_r}{2g}
            + H_4 \frac{\tau^{(n,m)}_\theta}{2g}\right)\, d\varphi
+\left(\frac{H_1}{r}\, dr +(1-H_2)\, d\theta \right)
  \frac{\tau^{(n)}_\varphi}{2g}
  , \label{a1} \end{equation}
\begin{equation}
A_\mu\, dx^\mu = \left( a_1\, dt + a_2\, \sin^2 \theta \, d\varphi \right)/g'
  , \end{equation}
and
\begin{equation}
\Phi = i\, 
      \left( \phi_1 \, \tau^{(n,m)}_r 
           + \phi_2  \tau^{(n,m)}_\theta \right)
    \frac{v}{\sqrt2} \left( \begin{array}{c} 0\\1  \end{array} \right) 
  . \end{equation}
where
\begin{eqnarray}          
\tau^{(n,m)}_r & = & \sin m\theta (\cos n\vphi \tau_x + \sin n\vphi \tau_y) 
           + \cos m\theta \tau_z \ , \ \ 
\nonumber \\       
\tau^{(n,m)}_\theta & = & \cos m\theta (\cos n\vphi \tau_x + \sin n\vphi \tau_y) 
           - \sin m\theta \tau_z \ , \ \ 
\nonumber \\       
\tau^{(n)}_\vphi & = & (-\sin n\vphi \tau_x + \cos n\vphi \tau_y) 
\ , \ \ \nonumber 
\end{eqnarray}
$n$ and $m$ are integers,
and $\tau_x$, $\tau_y$ and $\tau_z$ denote the Pauli matrices.

The two integers $n$ and $m$ determine the type of configuration,
that is put into rotation.
For $n=m=1$ the solutions correspond to rotating sphalerons.
Rotating multisphaleron configurations arise for $n>1$ and $m=1$.
For $n=1$ and $m>1$ rotating sphaleron-antisphaleron
pairs ($m=2$) or sphaleron-antisphaleron chains arise,
and for $n \ge 3$ rotating vortex ring solutions.

The ten functions $B_1$, $B_2$, $H_1,\dots,H_4$, $a_1$, $a_2$,
$\phi_1$, and $\phi_2$ depend on $r$ and $\theta$, only. 
With this Ansatz the full set of field equations reduces to a system 
of ten coupled partial differential equations in the independent variables 
$r$ and $\theta$. A residual $U(1)$ gauge degree of freedom is 
fixed by the condition $r\partial_r H_1 - \partial_\theta H_2=0$ 
\cite{Kleihaus:1991ks}.

Requiring regularity and finite energy, we impose 
for odd $m$ configurations the boundary conditions 
\begin{eqnarray}
r=0: &  & 
B_1 \sin m \theta + B_2 \cos m \theta =0  , \
\partial_r \left( B_1 \cos m \theta - B_2 \sin m \theta\right) =0  , \
H_1=H_3=H_4=0 , \ H_2=1 , \ 
\nonumber \\
&  & 
\partial_r a_1 =0  , \ a_2=0  , \
\phi_1=0 , \ \phi_2 = 0
\nonumber \\
r\rightarrow \infty: &  &  
B_1 = \gamma \cos m \theta  , \ B_2 = \gamma \sin m \theta  , \
H_1=H_3=0 , \ H_2=1-2m  ,  \ H_4=\frac{2\sin m\theta}{\sin\theta} , \ 
\nonumber \\[-0.7ex]
& & 
a_1=\gamma  , \ a_2 = 0  , \
\phi_1=1 , \ \phi_2 = 0 , \ 
 {\rm where } \ \gamma = const.
\nonumber \\[0.5ex]
\theta = 0: &  & 
\partial_\theta B_1 =0  , \ B_2 =0  , \
H_1=H_3=0 , \ \partial_\theta H_2=\partial_\theta H_4=0 , \ 
\partial_\theta a_1 = \partial_\theta a_2 = 0  , \
\partial_\theta \phi_1 = 0  , \ \phi_2=0  ,
\end{eqnarray}
where the latter hold also at $\theta=\pi/2$,
except for $B_1=0$ and $\partial_\theta B_2 =0$.
For even $m$ configurations the same set of boundary conditions
holds except for
\begin{eqnarray}
r=0: &  & 
\phi_1 \sin m \theta + \phi_2 \cos m \theta =0  , \
\partial_r \left( \phi_1 \cos m \theta - \phi_2 \sin m \theta\right) =0  
\nonumber \\
\theta=\pi/2: & &
\partial_\theta B_1=0 , \ B_2 = 0 \ , \partial_\theta H_3 = 0 , \ H_4=0 .
\end{eqnarray}

\section{Global charges of sphaleron-antisphaleron systems}

We now address the global charges of the sphaleron-antisphaleron systems,
their mass, their angular momentum, their electric charge, 
and their baryon number.
The mass $M$ and angular momentum $J$ are defined in terms of
volume integrals of the respective components
of the energy-momentum tensor. The mass is obtained from
\begin{equation}
M =- \int T_t^t d^3 r , 
\label{Mint}
\end{equation}
while the angular momentum
\begin{equation}
J = \int T_\varphi^t d^3 r 
= \int \left[ 2 {\rm Tr} \left( F^{t\mu} F_{\varphi\mu} \right)
 + f^{t\mu} f_{\varphi\mu} + 2 \left( D^t \Phi \right)^\dagger
 \left( D_\varphi \Phi \right)   \right] d^3 r
\label{Jint}
\end{equation}
can be reexpressed with help of the
equations of motion and the symmetry properties of the Ansatz
\cite{VanderBij:2001nm,Kleihaus:2002tc,Volkov:2003ew,Radu:2008pp}
as a surface integral at spatial infinity 
\begin{equation}
J =  \int_{S_2} \left\{ 
 2 {\rm Tr} \left( \left(V_\varphi - \frac{n\tau_z}{2g}\right) F^{r t} \right) 
 + 
 \left(A_\varphi - \frac{n}{g'}\right) f^{r t}
 \right\} r^2 \sin \theta d \theta d \varphi .
\label{Joint}
\end{equation}
The power law fall-off of the $U(1)$ field
of a charged solution allows for a finite flux
integral at infinity and thus a finite angular momentum.
Insertion of the asymptotic expansion for the $U(1)$ field
\begin{eqnarray}
a_1= \gamma - \frac{\chi}{r} + O \left(\frac{1}{r^2}\right) , \nonumber\\
a_2= \frac{\zeta}{r} + O \left(\frac{1}{r^2}\right) ,
\label{asymp}
\end{eqnarray}
and of the analogous expansion for the $SU(2)$ fields
then yields for the angular momentum
\begin{equation}
\frac{J}{4 \pi} = 
\frac{n\chi}{g^2} + \frac{n\chi}{{g'}^2}
= \frac{n\chi}{g^2 \sin^2 \theta_{\rm w}} = \frac{n\chi}{e^2} .
\label{Joint2}
\end{equation}

The field strength tensor ${\cal F}_{\mu\nu}$ of the 
electromagnetic field ${\cal A}_{\mu}$,
\begin{equation}
{\cal A}_{\mu}=\sin \theta_{\rm w} V^3_\mu + \cos \theta_{\rm w} A_\mu ,
\label{emfield}
\end{equation}
as given in a gauge where the Higgs field asymptotically tends to
Eq.~(\ref{Higgs}),
then defines the electric charge $\cal Q$
\begin{equation}
{\cal Q} =  \int_{S_2}
  {^*}{\cal F}_{\theta\varphi} d\theta d\varphi
 =4\pi\left\{ 
 \frac{\sin \theta_{\rm w} \chi}{g} + \frac{\cos \theta_{\rm w} \chi}{g'}
 \right\}
 = 4\pi\frac{\chi}{e}
  , \label{Qel} \end{equation}
where the integral is evaluated at spatial infinity.
%
Comparison of Eqs.~(\ref{Joint2}) and (\ref{Qel}) then
yields a linear relation between the angular momentum $J$
and the electric charge $\cal Q$ \cite{Radu:2008ta,Kleihaus:2008cv}
\begin{equation}
 J = \frac{n \cal Q}{e}
\label{JQrel}
 . \end{equation}
This relation 
corresponds to the relation for
monopole-antimonopole systems without magnetic charge 
\cite{Kleihaus:2007vf}.
The magnetic moment $\mu$ is obtained from 
the asymptotic expansion Eq.~(\ref{asymp}), analogously to the 
electric charge,
\begin{equation}
\mu= \frac{4\pi\zeta}{e} \ .
\end{equation}

Addressing finally the baryon number $Q_{\rm B}$,
its rate of change is given by
\begin{equation}
 \frac{d Q_{\rm B}}{dt} = \int d^3 r \partial_t j^0_{\rm B}
= \int d^3 r \left[ \vec \nabla \cdot \vec j_{\rm B}
 + \frac{1}{32 \pi^2} \epsilon^{\mu\nu\rho\sigma} \, \left\{
 g^2 {\rm Tr} \left(F_{\mu\nu} F_{\rho\sigma} \right) 
 + \frac{1}{2} {g'}^2 f_{\mu\nu} f_{\rho\sigma} \right\} \right]  . 
\end{equation}
Starting at time $t=-\infty$ at the vacuum with $Q_{\rm B}=0$,
one obtains the baryon number of a sphaleron solution at
time $t=t_0$ \cite{Klinkhamer:1984di},
\begin{equation}
 Q_{\rm B} = 
\int_{-\infty}^{t_0} dt \int_S \vec K \cdot d \vec S
+  \int_{t=t_0} d^3r K^0 
  , \end{equation}
where the $\vec \nabla \cdot \vec j_{\rm B}$ term is neglected,
and the anomaly term is reexpressed in terms of the
Chern-Simons current
\begin{equation}
 K^\mu=\frac{1}{16\pi^2}\varepsilon^{\mu\nu\rho\sigma} 
 \left\{ g^2 {\rm Tr}\left( F_{\nu\rho}V_\sigma
 - \frac{2}{3} i g V_\nu V_\rho V_\sigma \right)
 + \frac{1}{2} {g'}^2 f_{\nu\rho}A_\sigma \right\}
  . \end{equation}
In a gauge, where
\begin{equation}
V_\mu \to \frac{i}{g} \partial_\mu \hat{U} \hat{U}^\dagger   , \ \ \ 
\hat{U}(\infty) = 1   , 
\end{equation}
$\vec K$ vanishes at infinity. 
Subject to the above ansatz and boundary conditions
the baryon charge
of the sphaleron solution \cite{Kleihaus:1994tr,Kleihaus:2003tn} is then
\begin{equation}
Q_{\rm B}= \int_{t=t_0} d^3r K^0  = \frac{n \ (1-(-1)^m)}{4} \, .
\label{Q_B}
\end{equation}

\section{Results and discussion}


We have solved the set of ten coupled non-linear
elliptic partial differential equations numerically \cite{FIDISOL},
subject to the above boundary conditions
in compactified dimensionless coordinates,
$x = \tilde r/(1+ \tilde r)$, with $\tilde r = gvr$.
Restricting to $M_H=M_W$,
and employing the physical value for the mixing angle $\theta_{\rm w}$,
we have performed a systematic study
of the rotating sphaleron-antisphaleron systems
with $1 \le m \le 6$ and $1 \le n \le 6$.

Starting from a given static neutral solution
for a sphaleron-antisphaleron system characterized by
the integers $n$ and $m$,
we have constructed the corresponding branch of
rotating solutions, by slowly increasing
the asymptotic value of the parameter
$\tilde \gamma = \gamma/gv$,
which specifies the boundary conditions for the
time components of the gauge fields.
The rotating branch ends 
when the limiting value $\tilde \gamma_{\rm max}=1/2$
is reached \footnote{The
exponent arises as a combination from the mass term, 
obtained from the Higgs vaccuum expectation value,
and the non-Abelian gauge field interaction term,
and yields asymptotically an exponential decay with
decay constant proportional to $\sqrt{1-4\tilde \gamma^2}$.
}.
Here some of the gauge field functions
no longer decay exponentially.
This therefore precludes localized solutions 
for larger values of $\tilde \gamma$.


\subsection{Multisphalerons}

We first present our results
for the 
rotating multisphaleron solutions,
i.e., the branches of rotating solutions with $m=1$ and $n>1$.
We exhibit in Fig.~\ref{f-1}a
the asymptotic gauge field parameter
$\tilde \gamma = \gamma/gv$ versus the
scaled angular momentum of multisphaleron solutions
consisting of $n$ superposed sphalerons with $1\le n \le 6$.
Since the angular momentum $J$ increases with $n$,
we exhibit the angular momentum per sphaleron
$J/n$, choosing units of $J_0=4 \pi/g^2$.
As $\tilde \gamma$ increases from zero to 
its maximal value $\tilde \gamma_{\rm max}=1/2$,
the angular momentum increases monotonically.
Consequently the solutions have maximal spin
at $\tilde \gamma_{\rm max}$.

\begin{figure}[h!]
\lbfig{f-1}
\begin{center}
\hspace{0.0cm} (a)\hspace{-0.6cm}
\includegraphics[height=.25\textheight, angle =0]{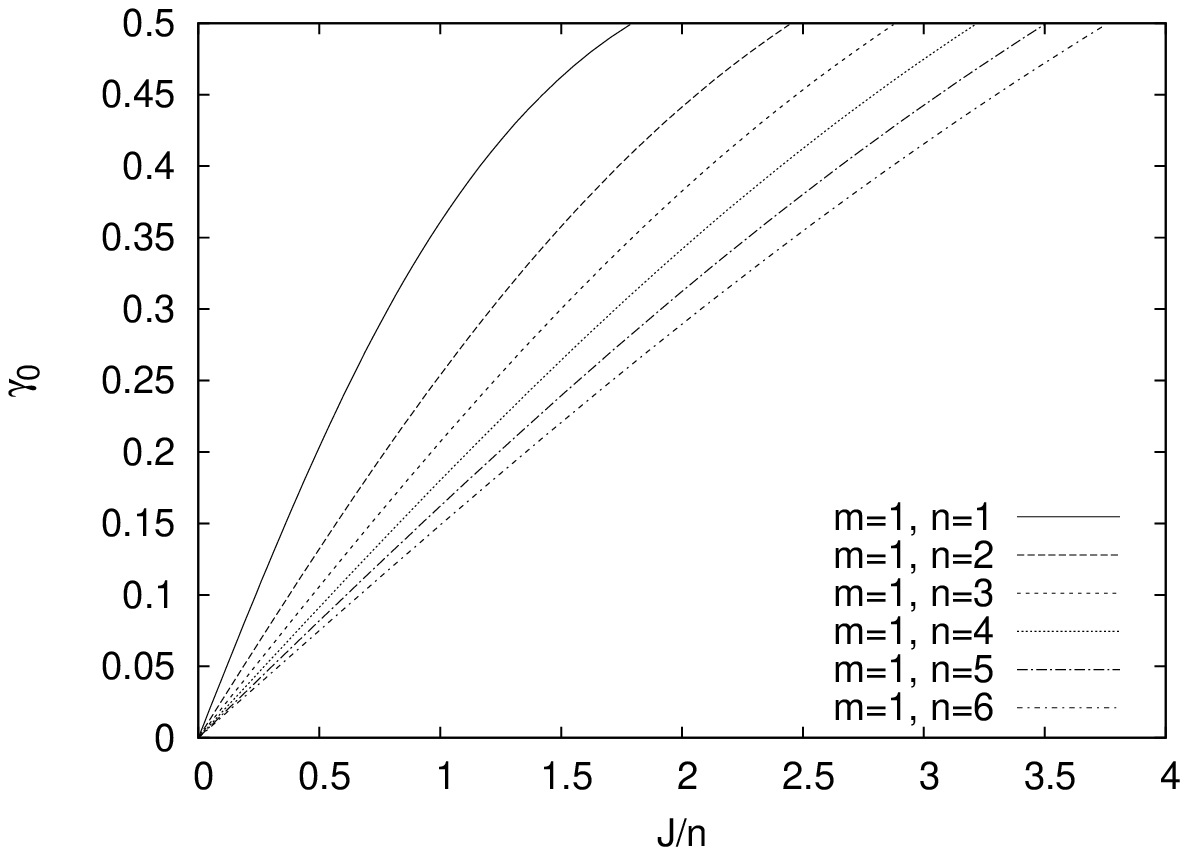}
\hspace{0.5cm} (b)\hspace{-0.6cm}
\includegraphics[height=.25\textheight, angle =0]{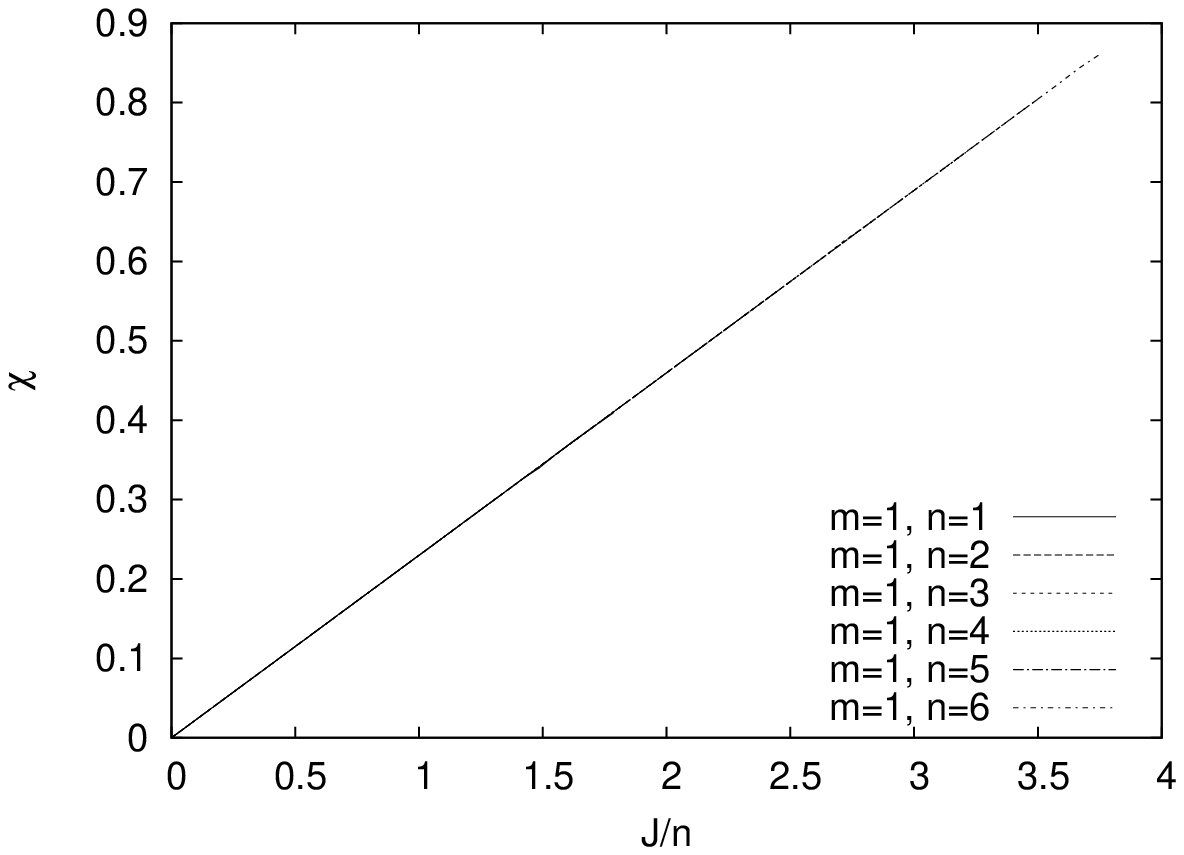}
\\
\hspace{0.0cm} (c)\hspace{-0.6cm}
\includegraphics[height=.25\textheight, angle =0]{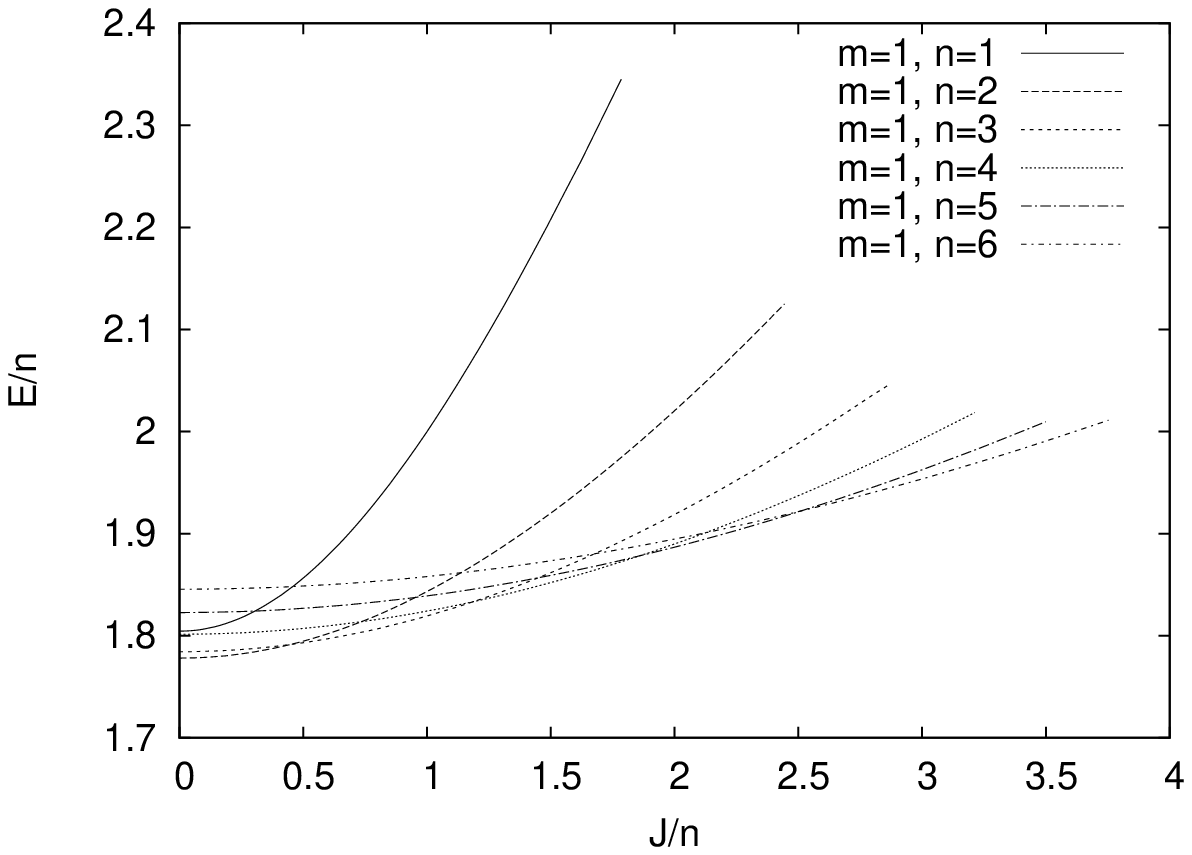}
\hspace{0.5cm} (d)\hspace{-0.6cm}
\includegraphics[height=.25\textheight, angle =0]{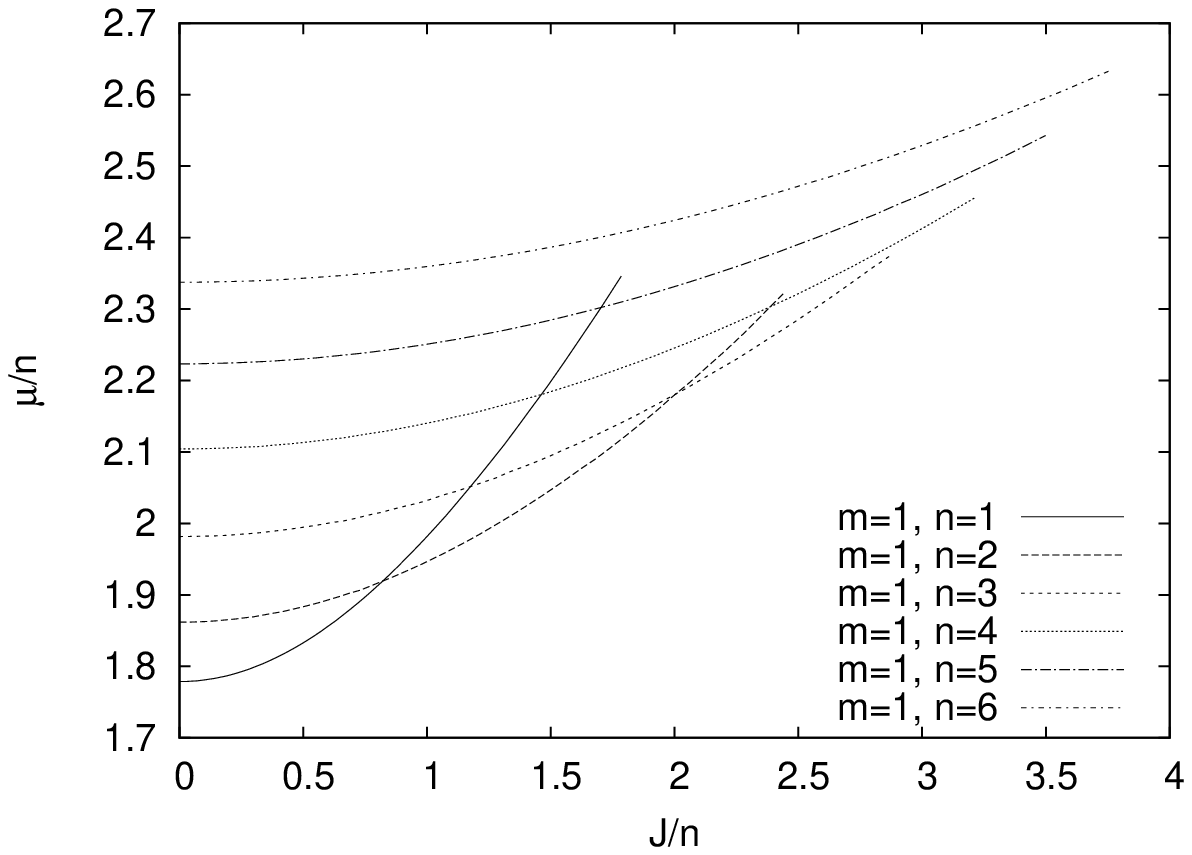}
\end{center}
\caption{
Properties of multisphaleron solutions ($m=1$, $n=1,\dots,6$):
(a) the asymptotic value of the $U(1)$ field $\tilde \gamma = \gamma/gv$,
(b) the $U(1)$ charge parameter $\chi$,
(c) the mass per sphaleron $E/n$ (in units of $E_0=4\pi v/g$),
(d) the magnetic moment per sphaleron $\mu/n$
(in units of $\mu_0=e/\alpha_{\rm W} M_{\rm W}$)
versus the angular momentum per sphaleron $J/n$
(in units of $J_0=4 \pi/g^2$)
}
\end{figure}

Considering the increase of the angular momentum with
the number of sphalerons $n$, 
we note that this increase is faster than linear. 
Thus the value of the maximal angular momentum 
per sphaleron, $J_{\rm max}/n$ increases with $n$. 
The more sphalerons a configuration
consists of, the more angular momentum the constituents
can contribute. 

Let us next consider the linear relation 
(\ref{JQrel}) between the charge $\cal Q$ and
the angular momentum $J$.
According to this relation,
we should obtain a single straight line, 
when exhibiting the charge 
versus the scaled angular momentum per sphaleron, $J/n$.
We demonstrate this in Fig.~\ref{f-1}b,
where we exhibit the charge parameter $\chi$
(which is proportional to the charge $\cal Q$)
versus the scaled angular momentum per sphaleron, $J/n$.
We indeed observe a single straight line,
which extends the further the greater $n$.
Since the charge parameter has been extracted from the
asymptotic fall-off of the $U(1)$ function $a_1$,
whereas the angular momentum has been obtained from
the volume integral of the angular momentum density $T^t_\varphi$,
this agreement reflects the good numerical quality of the
solutions.

In Fig.~\ref{f-1}c
we present the energy of these solutions,
which has been obtained from
the volume integral of the energy density $-T^t_t$.
For multisphalerons consisting of $n$ sphalerons one expects       
that their energy should be of the order of $n$ times
the energy of a single sphaleron,
thus $E/n$ should be roughly constant.
The deviations of the energy per sphaleron $E/n$
from the energy of the single sphaleron
can then be attributed to the interaction of the $n$ sphalerons
and be interpreted in terms of the binding energy
of these multisphaleron configurations.
For the employed value of the Higgs mass, we note that
the static solutions with $n=2-4$ represent bound states,
since $E/n$ is smaller than the energy of a single sphaleron,
whereas the static solutions with $n>4$ are slightly unbound
\cite{Kleihaus:1994tr}.

When charge is added to these configurations and the 
solutions begin to spin, their energy increases monotonically
with their angular momentum.
The increase of the energy per sphaleron $E/n$ 
with the angular momentum per sphaleron $J/n$ is strongest
for the branch of single sphaleron solutions.
The more sphalerons a configuration consists of,
the weaker is the increase of its energy per sphaleron $E/n$
with increasing angular momentum per sphaleron $J/n$.
Thus charge and rotation contribute relatively less
to the total energy for these ``many sphaleron'' configurations
(e.g.~only 8\% for $n=6$ as compared to 30\% for $n=1$).
Consequently, the rotating multisphaleron configurations
turn into bound states beyond some critical value
of the angular momentum.

Sphalerons possess a large magnetic moment $\mu$.
For multisphalerons consisting of $n$ sphalerons one 
then expects 
that their magnetic moment should be roughly $n$ times
the magnetic moment $\mu$ of a single sphaleron.
As seen in Fig.~\ref{f-1}d, where we exhibit 
the magnetic moment per sphaleron $\mu/n$ versus
the angular momentum per sphaleron $J/n$,
this first guess is somewhat crude.
For static configurations,
the interaction between the sphalerons 
gives rise to a systematic (almost linear) increase of 
the magnetic moment per sphaleron $\mu/n$ with
the number of sphalerons.

When charge and thus angular momentum
is added to these configurations, 
their magnetic moment increases monotonically
with increasing angular momentum.
Again, this increase is strongest
for the branch of single sphaleron solutions,
and the increase is the weaker 
the more sphalerons a configuration consists of.

Having discussed the global properties of these solutions,
we now turn to their local properties.
In particular, we address the effect of 
the presence of charge and rotation
on the energy density $-T^t_t$,
and on the modulus of the Higgs field $|\Phi|$.
As an example, we illustrate the energy density $-T^t_t$,
the magnitude of the Higgs field $|\Phi|$,
and the angular momentum density $T^t_\varphi$
for an almost maximally rotating
multisphaleron solution ($m=1$, $n=3$, $\tilde \gamma =0.499$)
in Fig.~\ref{f-2}.

\begin{figure}[h!]
\lbfig{f-2}
\begin{center}
\hspace{0.0cm} (a)\hspace{-0.6cm}
\includegraphics[height=.25\textheight, angle =0]{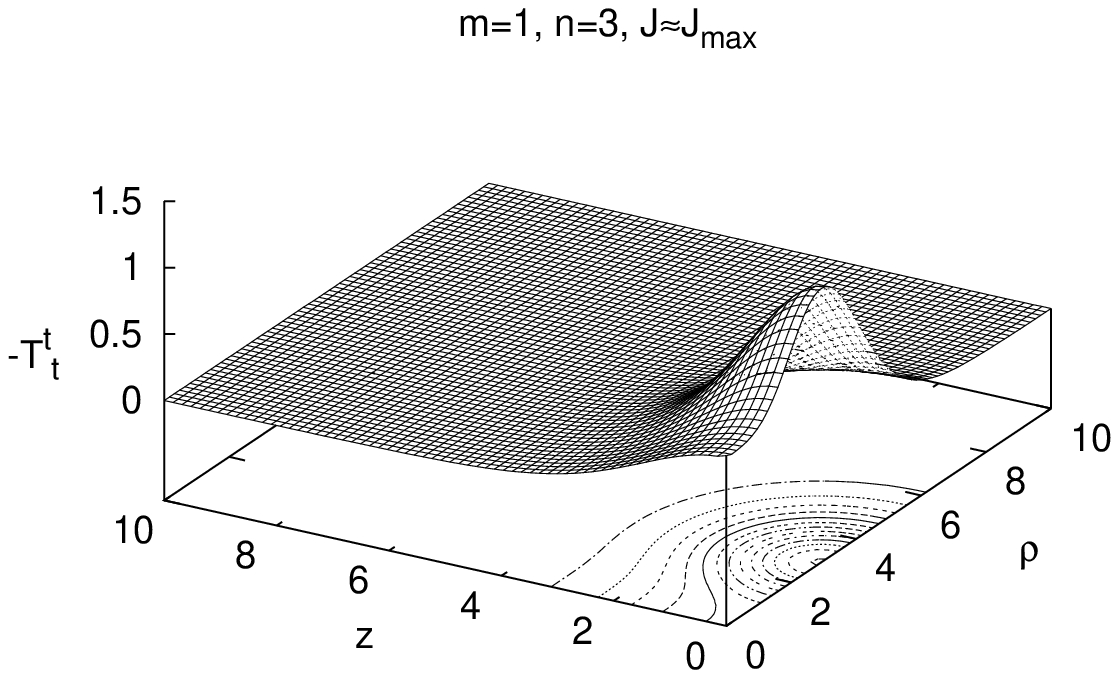}
\hspace{0.5cm} (b)\hspace{-0.6cm}
\includegraphics[height=.25\textheight, angle =0]{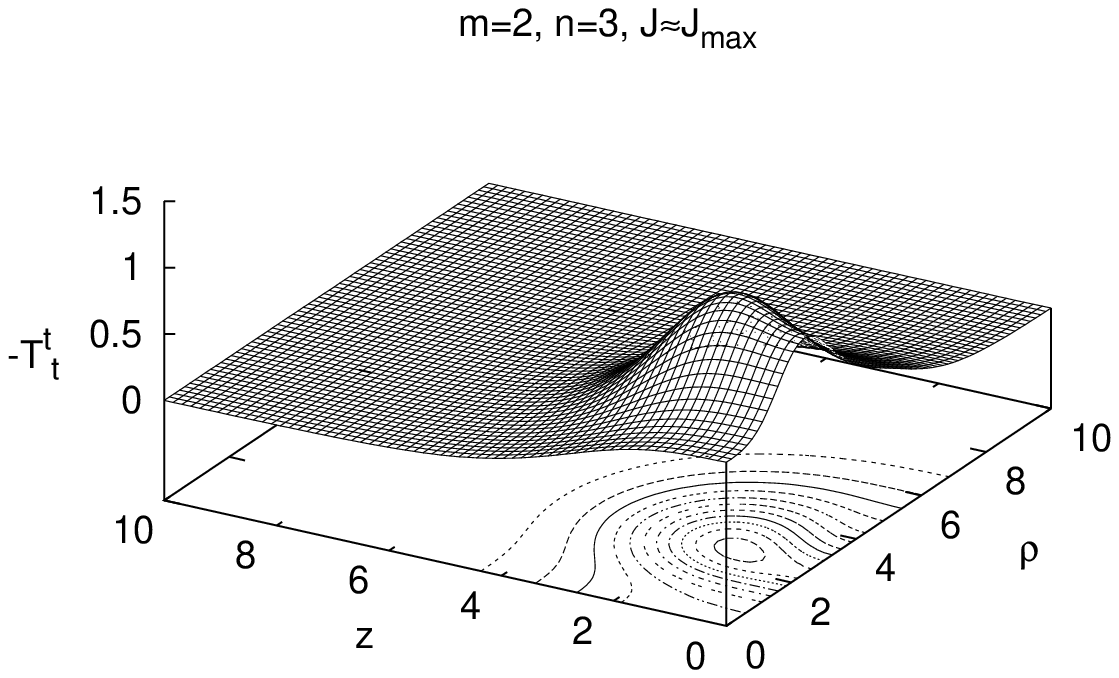}
\\
\hspace{0.0cm} (c)\hspace{-0.6cm}
\includegraphics[height=.25\textheight, angle =0]{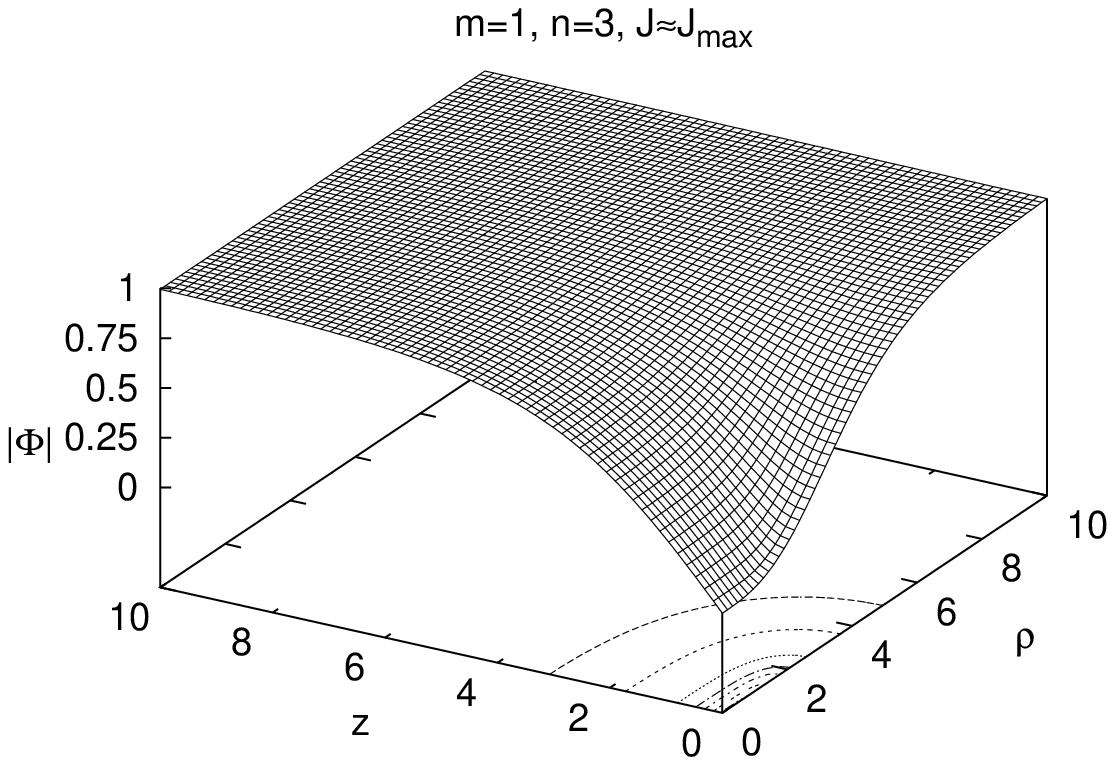}
\hspace{0.5cm} (d)\hspace{-0.6cm}
\includegraphics[height=.25\textheight, angle =0]{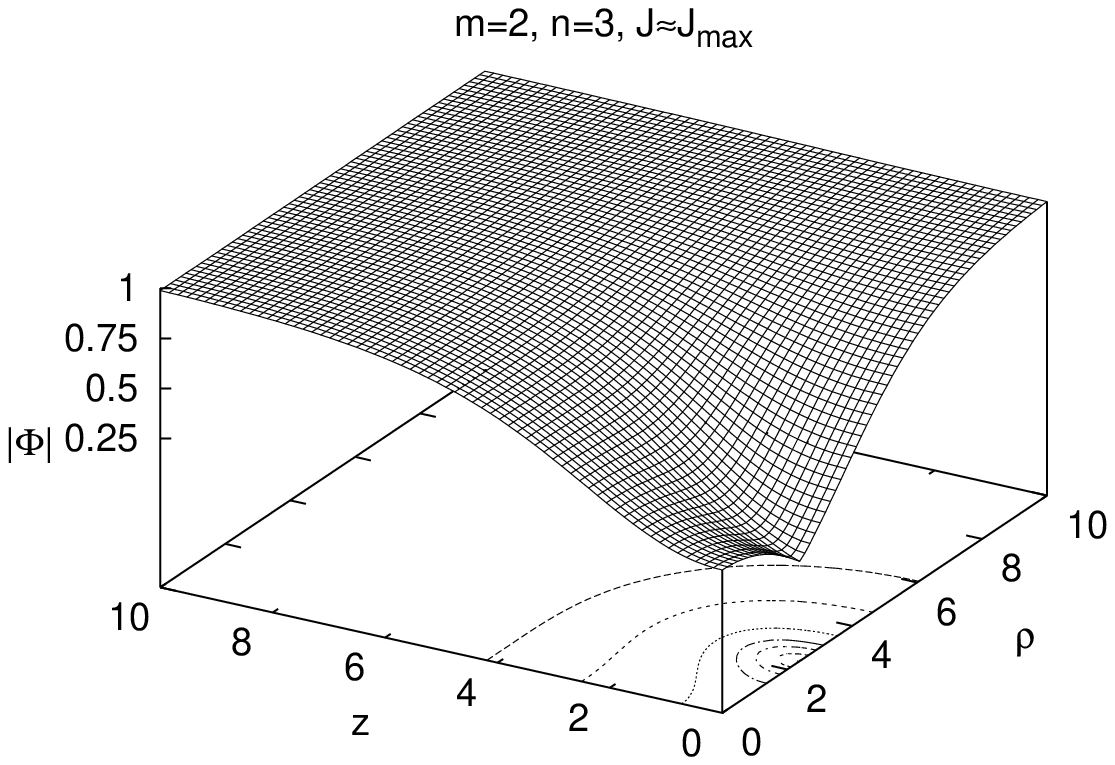}
\\
\hspace{0.0cm} (e)\hspace{-0.6cm}
\includegraphics[height=.25\textheight, angle =0]{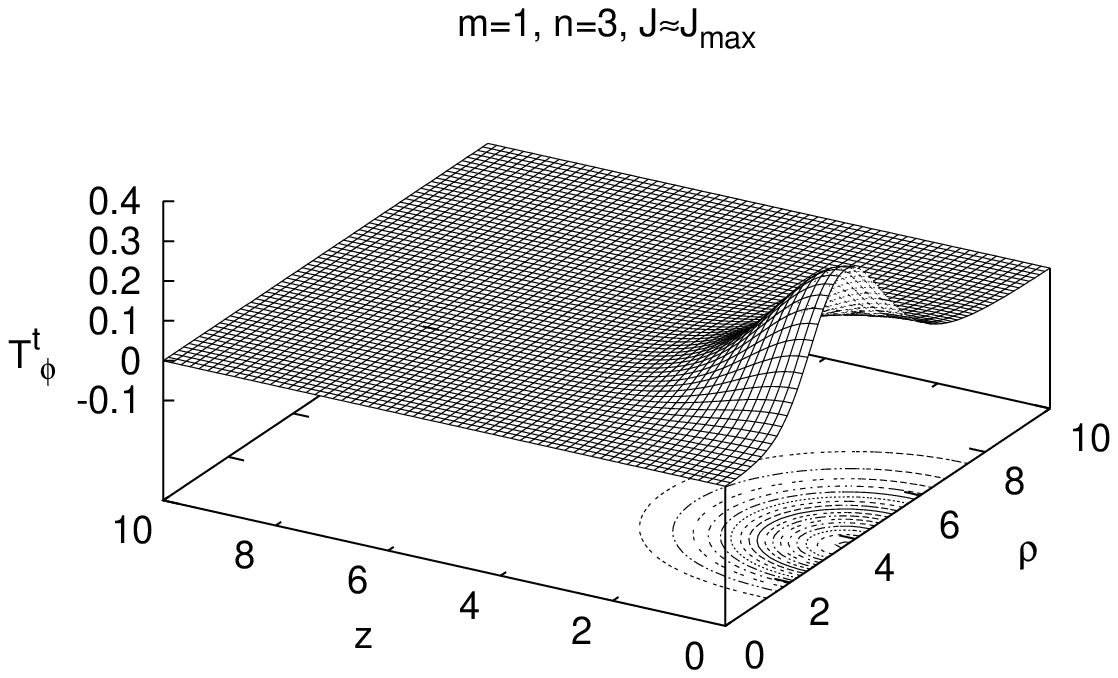}
\hspace{0.5cm} (f)\hspace{-0.6cm}
\includegraphics[height=.25\textheight, angle =0]{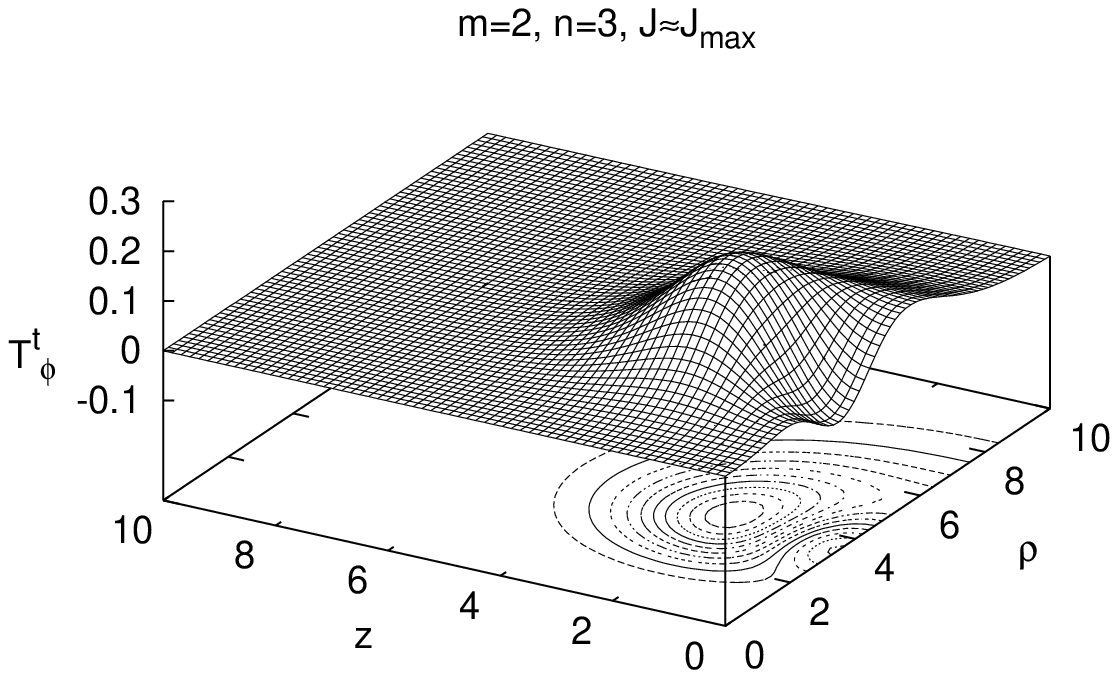}
\end{center}
\vspace{-0.5cm}
\caption{\small
The energy density $-T^t_t$ (top),
the modulus of the Higgs field $|\Phi|$ (middle),
and the angular momentum density $T^t_\varphi$ (bottom)
are exhibited for an $m=1$, $n=3$ multisphaleron solution (left)
and an $m=2$, $n=3$ sphaleron-antisphaleron system (right)
close to maximal rotation.
}
\end{figure}

In multisphalerons, the region with large energy density 
is torus-like, where its maximum is forming a ring in 
the equatorial plane.
When we add electric charge and angular momentum
to the static configuration, 
we observe that 
the energy density is spreading further out,
while at the same time its overall 
magnitude is reduced.
Such a spreading of the energy density with increasing charge is
also seen in dyons, for instance,
and we attribute this effect to the presence of charge
and the associated repulsion.
Indeed, this spreading
becomes quite pronounced for large values of the charge.
The expected effect of the presence of angular momentum,
on the other hand,
is a centrifugal shift of the energy density. 
Indeed, we observe, that with increasing angular momentum
the torus-like region of large energy density
moves further outward to larger values of $\rho$.

The modulus of the Higgs field of the multisphaleron
solutions has a single node at the origin,
from where it starts to increase linearly 
in the direction of the symmetry axis,
to reach its vacuum expectation value at infinity.
In the equatorial plane, in constrast,
the modulus of the Higgs field starts to
inrease from the origin much slower (i.e., only with power $\rho^n$).
As the configurations are endowed with
charge and rotation, the Higgs field hardly changes.
Indeed, the effect of charge and rotation on
the modulus of the Higgs field is barely noticable,
even at the maximal strength.

The angular momentum density for the multisphaleron solutions
is torus-like and centered in
the equatorial plane analogous to the energy density.
However, the region of large angular momentum density
is located further outwards at larger values of $\rho$,
while it vanishes on the symmetry axis.

\subsection{Sphaleron-antisphaleron systems}

Let us now turn to sphaleron-antisphaleron systems.
For $n=1$ they represent sphaleron-antisphaleron chains,
where $m$ sphalerons and antisphalerons are located
on the symmetry axis, in static equilibrium.
For $n=2$ the chain is formed
from  $m$ multisphalerons and -antisphalerons,
thus the modulus of the Higgs field
still possesses only isolated nodes on the symmetry axis.
However, as $n$ increases further, the character of the solutions
changes, and new types of configurations appear,
where the modulus of the Higgs field vanishes on rings
centered around the symmetry axis.
Therefore we refer to these solutions as vortex ring solutions.
We note, that the precise evolution of the isolated nodes 
on the symmetry axis and the vortex rings in the bulk
with increasing $n$ is sensitive
to the value of the Higgs mass \cite{Kleihaus:2008gn}.

In the following we demonstrate the effect of charge
and rotation on these sphaleron-antisphaleron systems
by focussing on the configurations with $n=3$ and
$m=1,...,6$. 
The static solutions have been constructed before \cite{Kleihaus:2008gn}.
For the chosen parameters, the modulus of the Higgs field
of the static $m=2$ configuration
vanishes only on a ring in the equatorial plane.
The static $m=3$ configuration 
has a node at the origin and in addition two rings,
located symmetrically above and below the $xy$-plane.
The static $m=4$ configuration has only two symmetrical vortex rings.
For $m=5$, the two symmetrical vortex rings
are supplemented by a node at the origin,
and for $m=6$ two symmetrical vortex rings and two inner nodes 
on the symmetry axis are present.

Again we first discuss the global properties of these solutions,
following the above scheme.
Since only the integer $n$ (but not $m$) enters the
relation between the charge and the angular momentum,
we continue to show all quantities versus 
the scaled relative angular momentum $J/n$.
We exhibit in Fig.~\ref{f-3}a
the dependence of the asymptotic gauge field parameter
$\tilde \gamma = \gamma/gv$ on the
scaled relative angular momentum $J/n$
for the configurations with $n=3$ and $m=1,...,6$.
As before, at the maximal value $\tilde \gamma_{\rm max}=1/2$
the solutions have maximal angular momentum.
Moreover, we observe that
the maximal value of the relative angular momentum
$J_{\rm max}/n$ increases with $m$.
Thus the higher the number of constituents a configuration
(i.e., the product $mn$),
the more angular momentum each of the constituents
can contribute.

\begin{figure}[h!]
\lbfig{f-3}
\begin{center}
\hspace{0.0cm} (a)\hspace{-0.6cm}
\includegraphics[height=.25\textheight, angle =0]{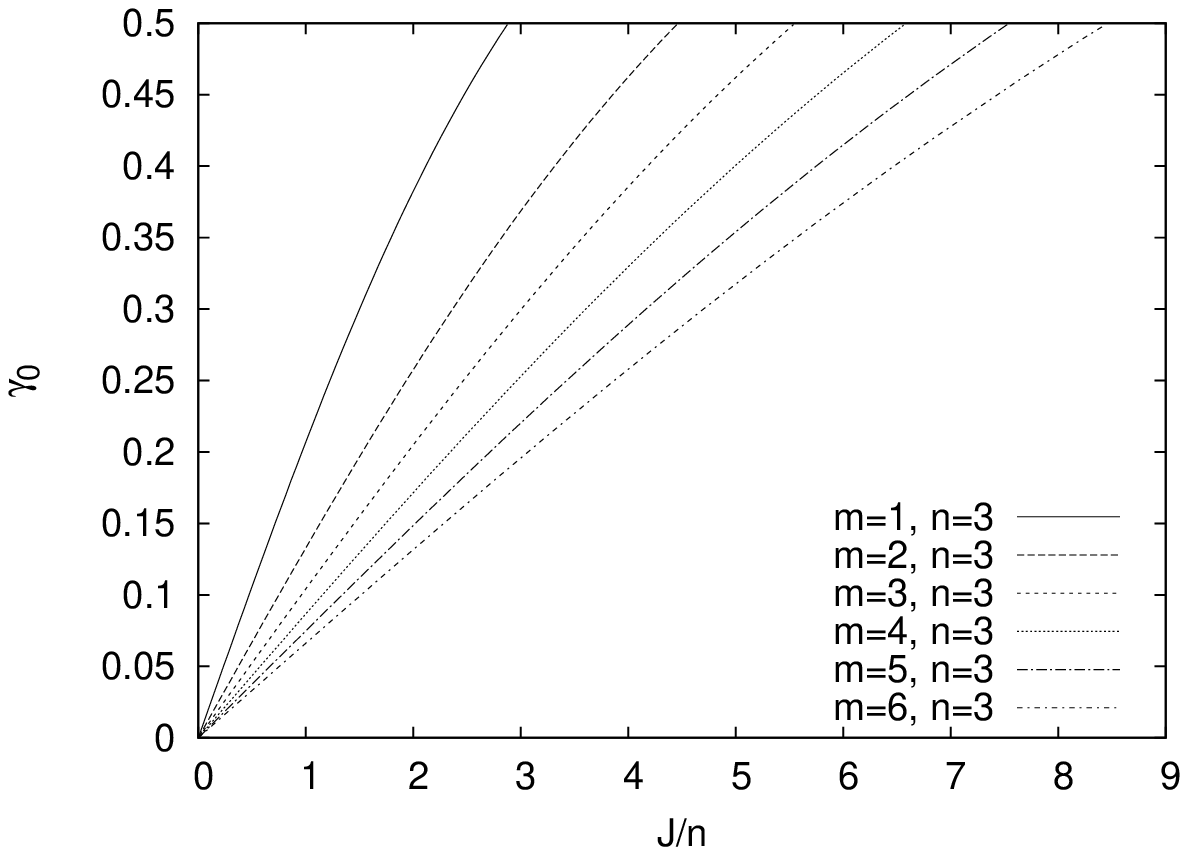}
\hspace{0.5cm} (b)\hspace{-0.6cm}
\includegraphics[height=.25\textheight, angle =0]{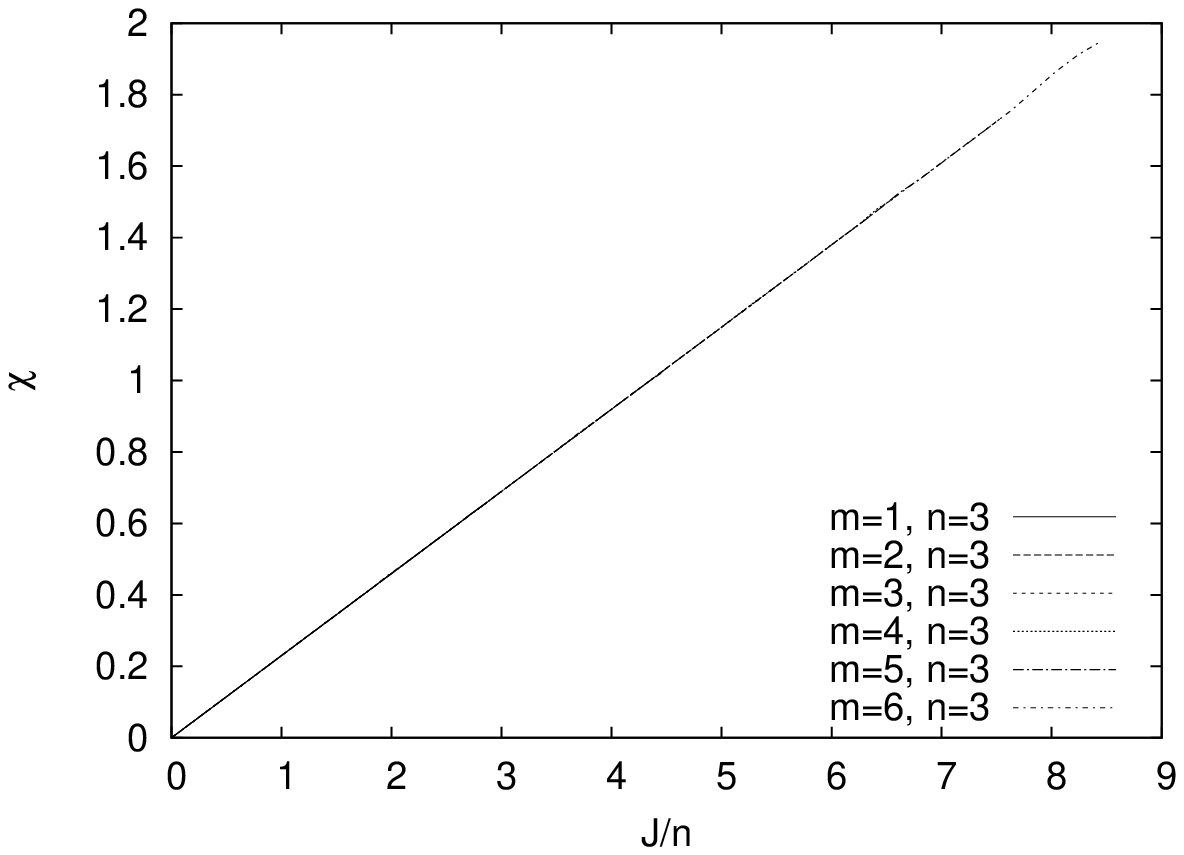}
\\
\hspace{0.0cm} (c)\hspace{-0.6cm}
\includegraphics[height=.25\textheight, angle =0]{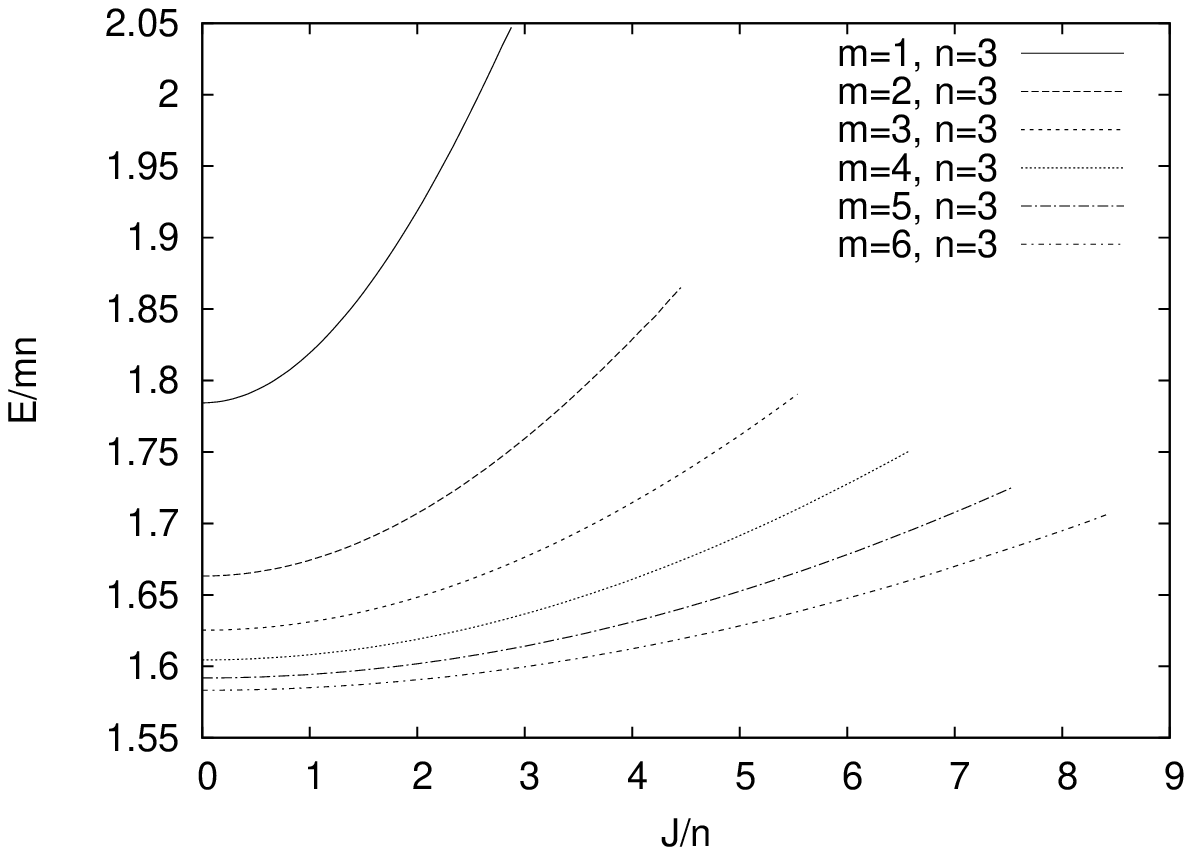}
\hspace{0.5cm} (d)\hspace{-0.6cm}
\includegraphics[height=.25\textheight, angle =0]{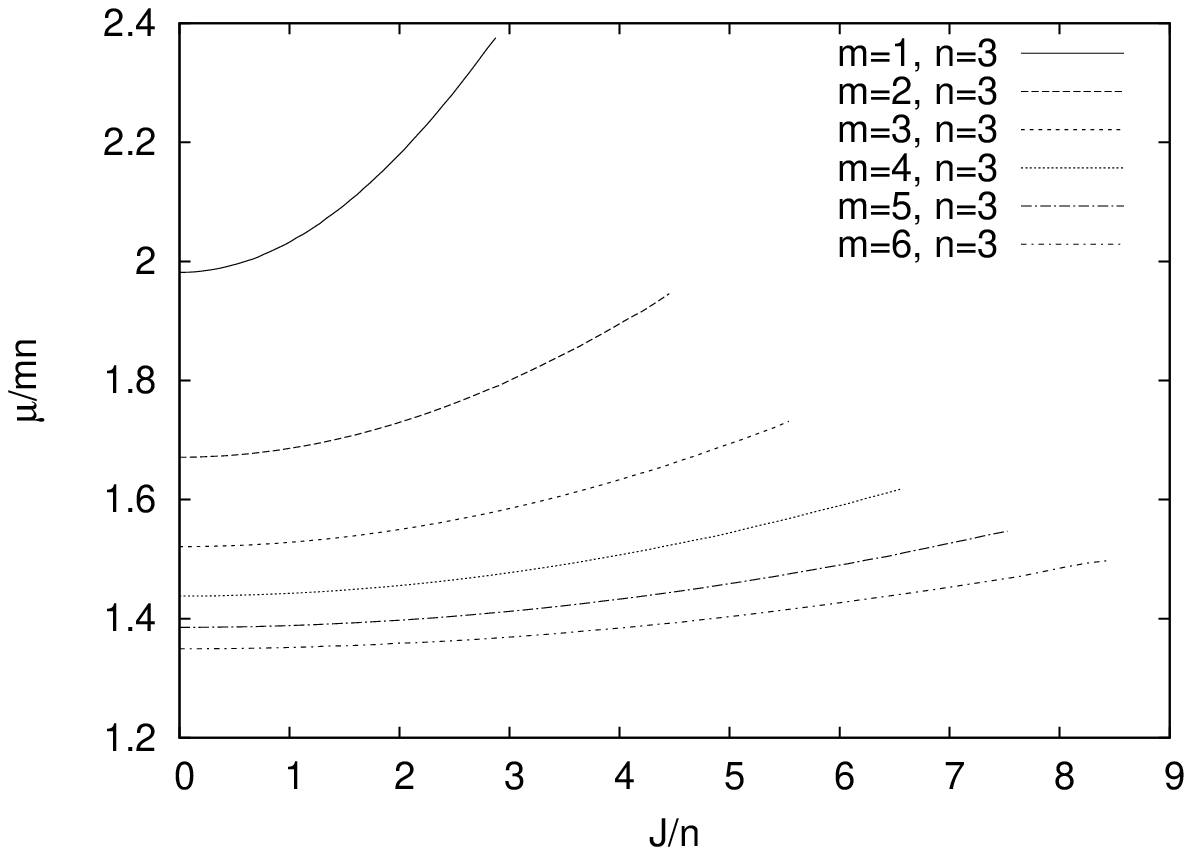}
\end{center}
\caption{
Same as Fig.\ref{f-1}
for sphaleron-antisphaleron systems ($m=1,\dots,6$, $n=3$).
}
\end{figure}

In Fig.~\ref{f-3}b we demonstrate
that the linear relation (\ref{JQrel}) 
between the charge $\cal Q$ and the angular momentum $J$
is well satisfied also by the numerically constructed
sphaleron-antisphaleron systems.
We again obtain a single straight line, 
when exhibiting the charge 
versus the relative angular momentum $J/n$,
which extends the further the greater $n$.

Addressing the energy of these solutions,
we expect that it should be roughly proportional
to the number $mn$ of constituents of the configurations.
We therefore consider the relative energy $E/mn$,
i.e., the energy per constituent.
The deviations of $E/mn$
from the energy of a single sphaleron
can then be attributed to the interaction of the sphalerons
and antisphalerons in the system
and be interpreted in terms of their binding energy.
We present the relative energy $E/mn$ of these solutions
in Fig.~\ref{f-3}c.
We note that the binding energy increases with an
increasing number of constituents.
Also, the increase of the energy per constituent $E/mn$ 
with $J/n$ is the stronger the smaller the number of constituents.
Charge and rotation contribute therefore relatively less
to the total energy for the ``many constituents'' configurations.

Addressing finally the magnetic moment $\mu$ of these
solutions, we also consider
the magnetic moment per constituent $\mu/mn$.
As seen in Fig.~\ref{f-3}d, 
the interaction between the constituents
gives rise to a decrease of
the magnetic moment per constituent $\mu/mn$ with
increasing $m$.

Now we turn to the local properties of these 
sphaleron-antisphaleron systems.
As examples, we exhibit the $m=2$, $n=3$
system in Fig.~\ref{f-2} (right column)
and the $m=3$, $n=3$ and $m=4$, $n=3$ systems in Fig.~\ref{f-4},
where we again display the energy density,
the modulus of the Higgs field and
the angular momentum density 
for solutions close to maximal rotation.

\begin{figure}[h!]
\lbfig{f-4}
\begin{center}
\hspace{0.0cm} (a)\hspace{-0.6cm}
\includegraphics[height=.25\textheight, angle =0]{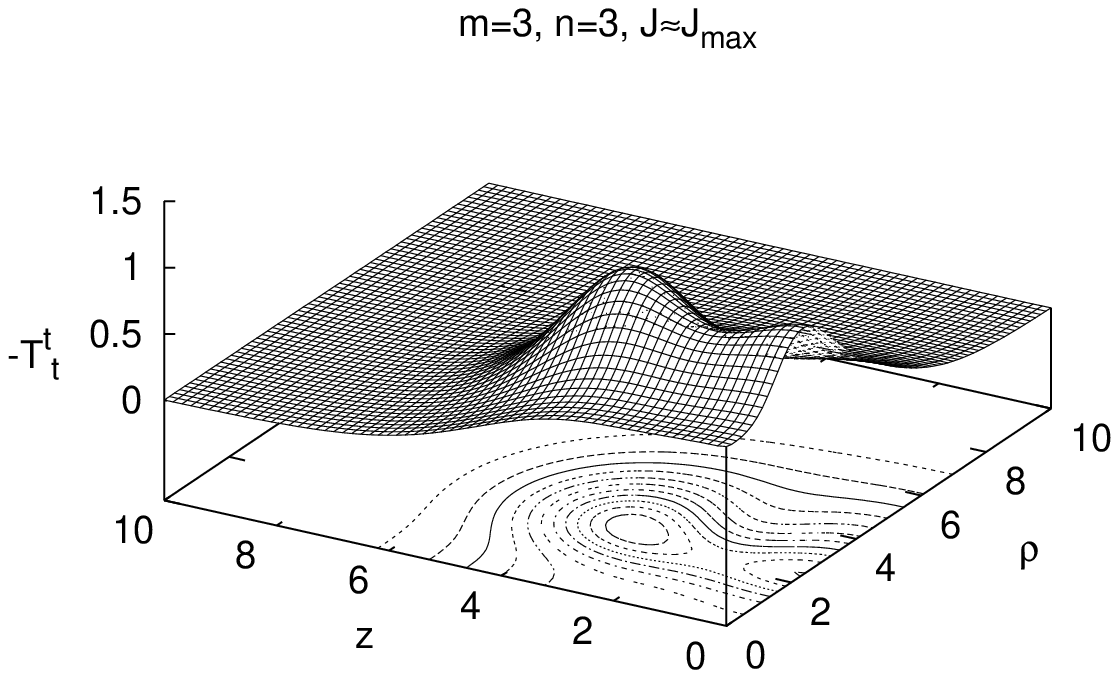}
\hspace{0.5cm} (b)\hspace{-0.6cm}
\includegraphics[height=.25\textheight, angle =0]{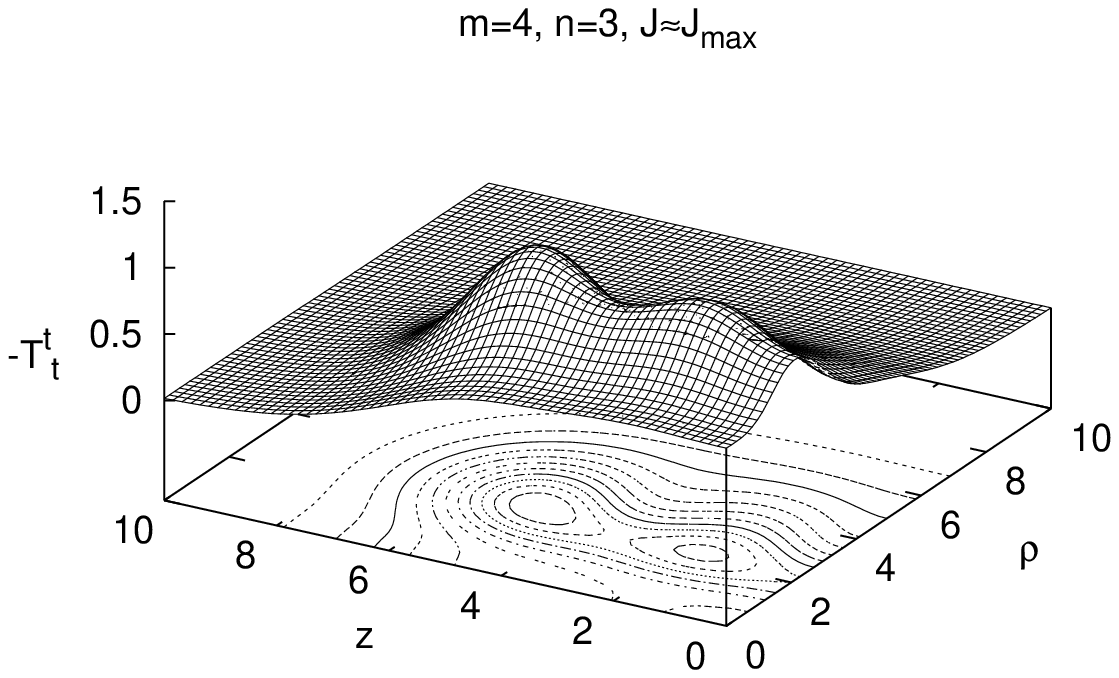}
\\
\hspace{0.0cm} (c)\hspace{-0.6cm}
\includegraphics[height=.25\textheight, angle =0]{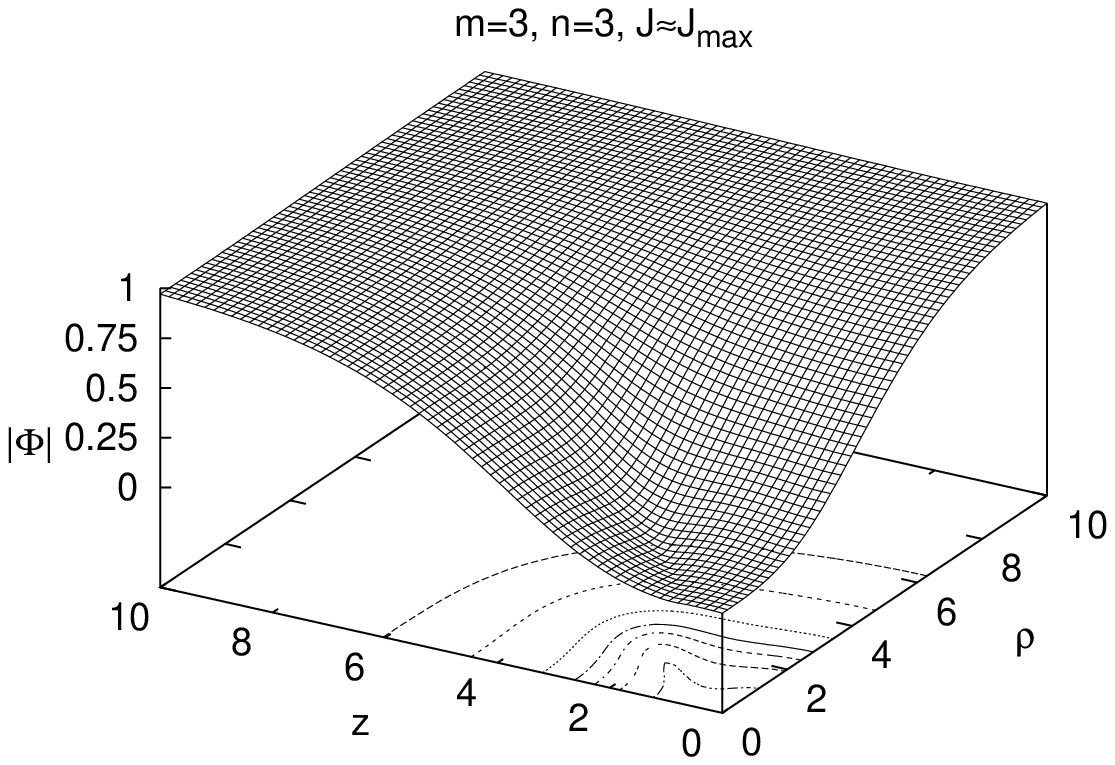}
\hspace{0.5cm} (d)\hspace{-0.6cm}
\includegraphics[height=.25\textheight, angle =0]{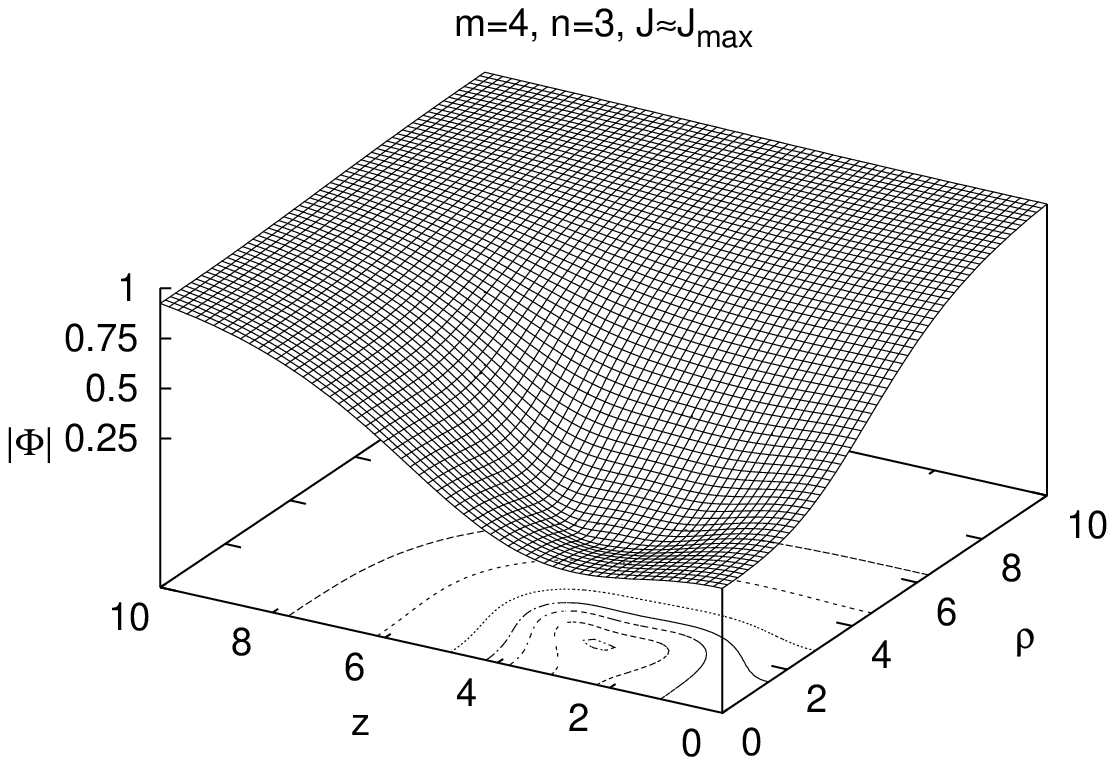}
\\
\hspace{0.0cm} (e)\hspace{-0.6cm}
\includegraphics[height=.25\textheight, angle =0]{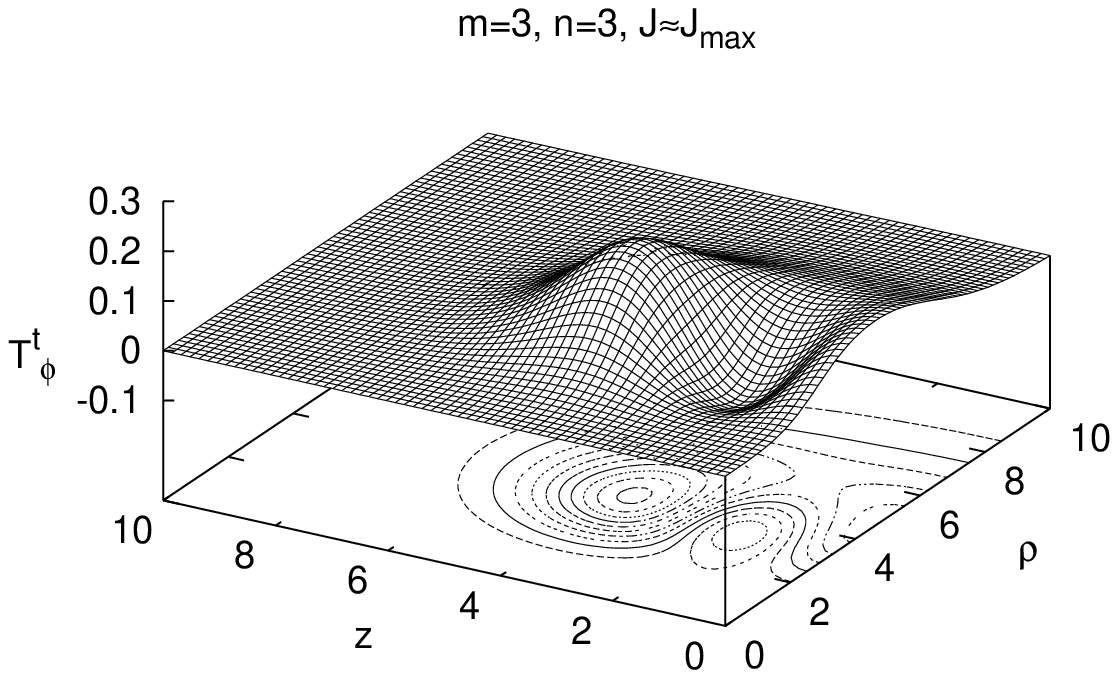}
\hspace{0.5cm} (f)\hspace{-0.6cm}
\includegraphics[height=.25\textheight, angle =0]{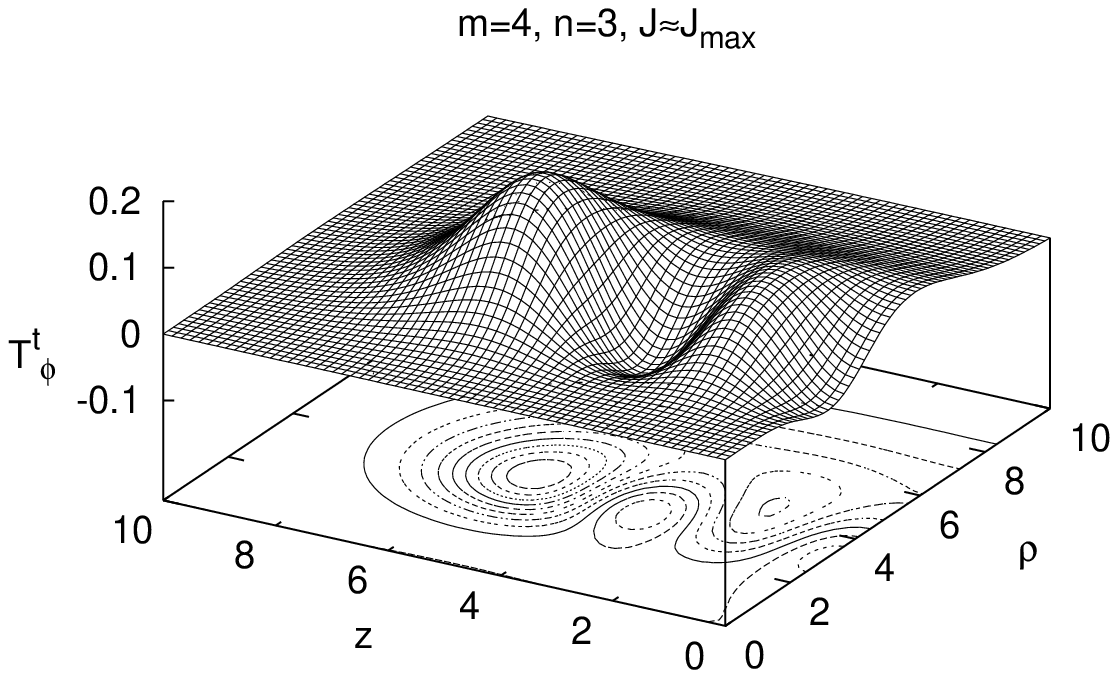}
\end{center}
\vspace{-0.5cm}
\caption{\small
Same as Fig.\ref{f-2}
for the sphaleron-antisphaleron systems $m=3$, $n=3$ (left)
and $m=4$, $n=3$ (right).
}
\end{figure}

Only in sphaleron-antisphaleron pairs and chains with $n=1$
the regions with large energy density 
are localized around the Higgs field nodes 
on the symmetry axis.
In all other sphaleron-antisphaleron systems, 
the regions with large energy density are torus-like,
where a configuration may possess several such tori.
The location and number of these tori
depends on the numbers $m$ and $n$ and on the parameters,
just as the nodes of the Higgs field are
determined by these quantities.

As for the multisphalerons,
the effect of the presence of electric charge
is that the energy density spreads further out,
while at the same time its overall 
magnitude reduces.
Likewise, the effect of the presence of angular momentum
is a centrifugal shift of the energy density. 
With increasing angular momentum
the torus-like regions of large energy density
move further outward to larger values of $\rho$.

The modulus of the Higgs field of the 
sphaleron-antisphaleron systems changes very little
and is barely noticable,
even when the systems carry maximal charge
and angular momentum.

The angular momentum density of the sphaleron-antisphaleron systems
is also characterized by the presence of tori.
But it contains tori of large positive angular momentum density
as well as negative angular momentum density.
The tori of the angular momentum density are spatially related to the tori
of the energy density.
In particular, the location of the positive tori is associated with
the location of the tori of the energy density,
with the negative tori inbetween.

\subsection{Equilibrium condition}

Finally, we would like to address the question of the
equilibrium of such composite configurations
as sphaleron-antisphaleron pairs and more general
sphaleron-antisphaleron systems.
As discussed previously \cite{Aharonov:1992wf,Beig:2008qi,Beig:2009jd},
a necessary condition for the equilibrium
of such axially symmetric configurations is 
\begin{equation}
\int_S T_{zz} dS =0
\label{Tzz}
\end{equation}
where $T_{zz}$ is the respective component
of the stress energy tensor and $S$ is the equatorial plane.
When this condition is satisfied, the net force between
the constituents in the upper and in the lower
hemisphere vanishes, thus yielding equilibrium.
If $T_{zz}$ vanishes everywhere in the equatorial plane,
this condition is met trivially,
if on the other hand $T_{zz}$ does not identically vanish, 
the various contributions to the surface integral 
(\ref{Tzz}) must precisely cancel each other.

To understand how the equilibrium condition is satisfied
in these sphaleron-antisphaleron systems,
we have extracted the $T_{zz}$ component of the
stress energy tensor. 
We illustrate 
$T_{zz}$ for two rather different configurations in Fig.~\ref{f-5}.
In Fig.~\ref{f-5}a 
we display $T_{zz}$
for the static sphaleron-antisphaleron pair 
in the upper hemisphere. 
In the equatorial plane $T_{zz}$ appears to almost vanish.
We therefore focus on the equatorial plane in Fig.~\ref{f-5}c. 
Here $T_{zz}$ is small, but finite (except when it changes sign).
To gain further insight into how the equilibrium 
results from the various forces present in the system,
we consider the contributions from the respective parts of the Lagrangian
separately.
We exhibit these also in Fig.~\ref{f-5}c. 
We note, that
the positive contribution from the $SU(2)$ gauge field part
almost cancels the negative contributions from the $U(1)$ and Higgs
parts, yielding in total a $T_{zz}$ which is almost but not quite vanishing
in the equatorial plane.
In the inner region the total $T_{zz}$ is slightly positive, while in the
outer region it is slightly negative, yielding
together a vanishing surface integral (\ref{Tzz}),
within the numerical accuracy.

The situation is similar for the sphaleron-antisphaleron
chain, consisting of four constituents ($m=4$, $n=1$).
Also, the inclusion of rotation does not change this
overall behaviour of these types of solutions.
For most other systems, however, $T_{zz}$ does not nearly vanish
in the equatorial plane. 
This is exhibited examplarily in Fig.~\ref{f-5}b
for the fast rotating sphaleron-antisphaleron system with
$m=4$, $n=3$. The features of $T_{zz}$
seen here, are very typical, and hardly change with rotation,
since the effect of rotation is basically a slight shift in magnitude.
However, while $T_{zz}$ is rather large in the equatorial plane
for these configurations,
its positive and negative contributions to the surface integral 
do cancel as required for equilibrium.

\begin{figure}[h!]
\lbfig{f-5}
\begin{center}
\hspace{0.0cm} (a)\hspace{-0.6cm}
\includegraphics[height=.25\textheight, angle =0]{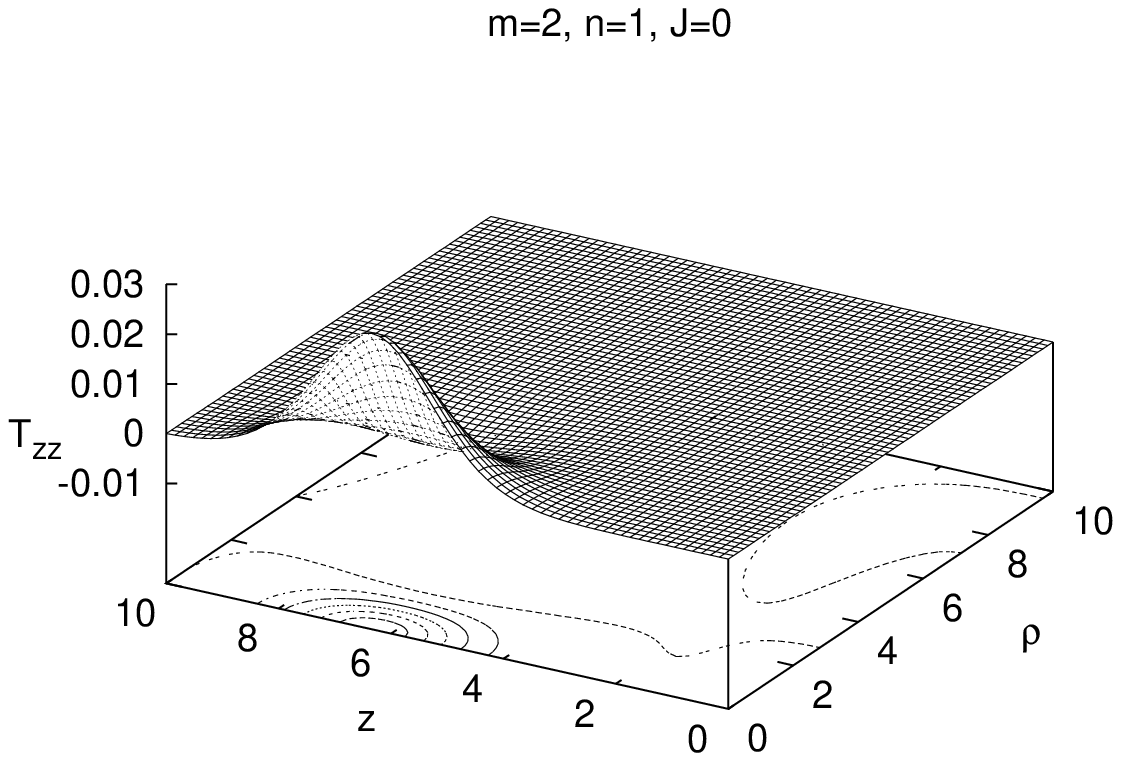}
\hspace{0.5cm} (b)\hspace{-0.6cm}
\includegraphics[height=.25\textheight, angle =0]{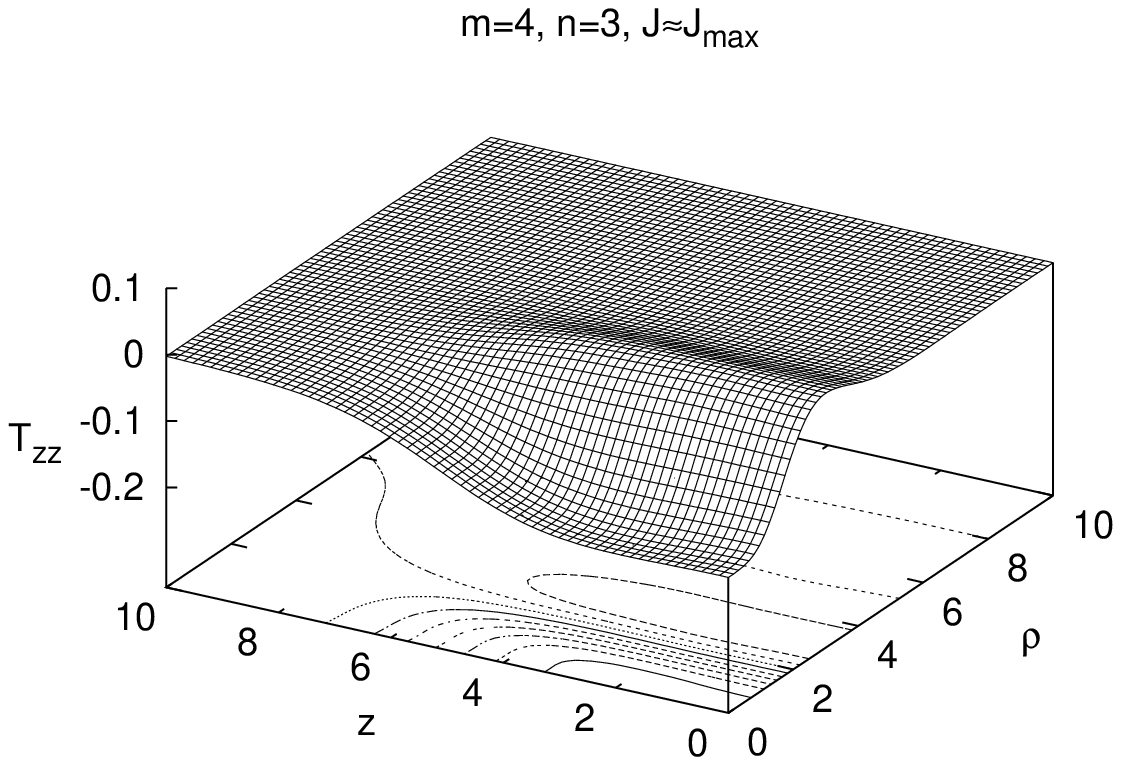}
\\
\hspace{0.0cm} (c)\hspace{-0.6cm}
\includegraphics[height=.23\textheight, angle =0]{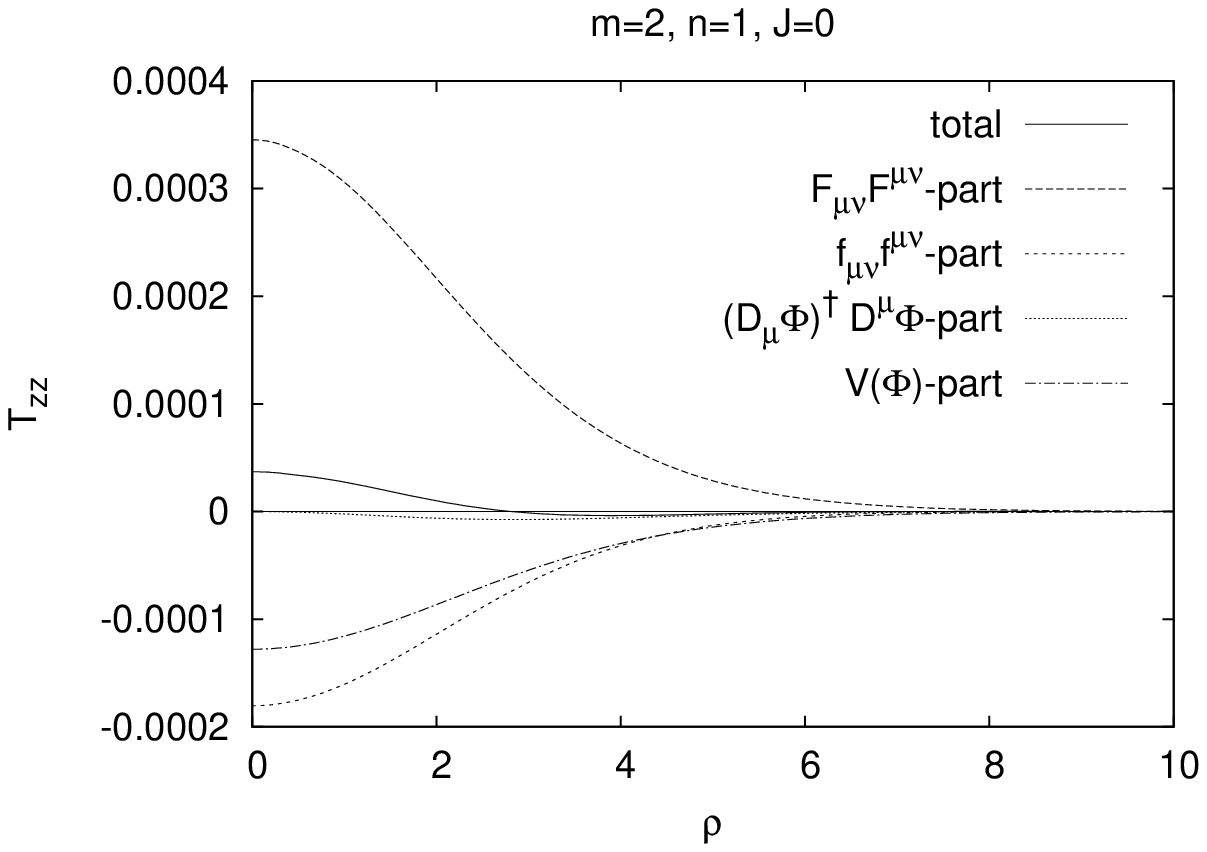}
\hspace{1.0cm} (d)\hspace{-0.6cm}
\includegraphics[height=.23\textheight, angle =0]{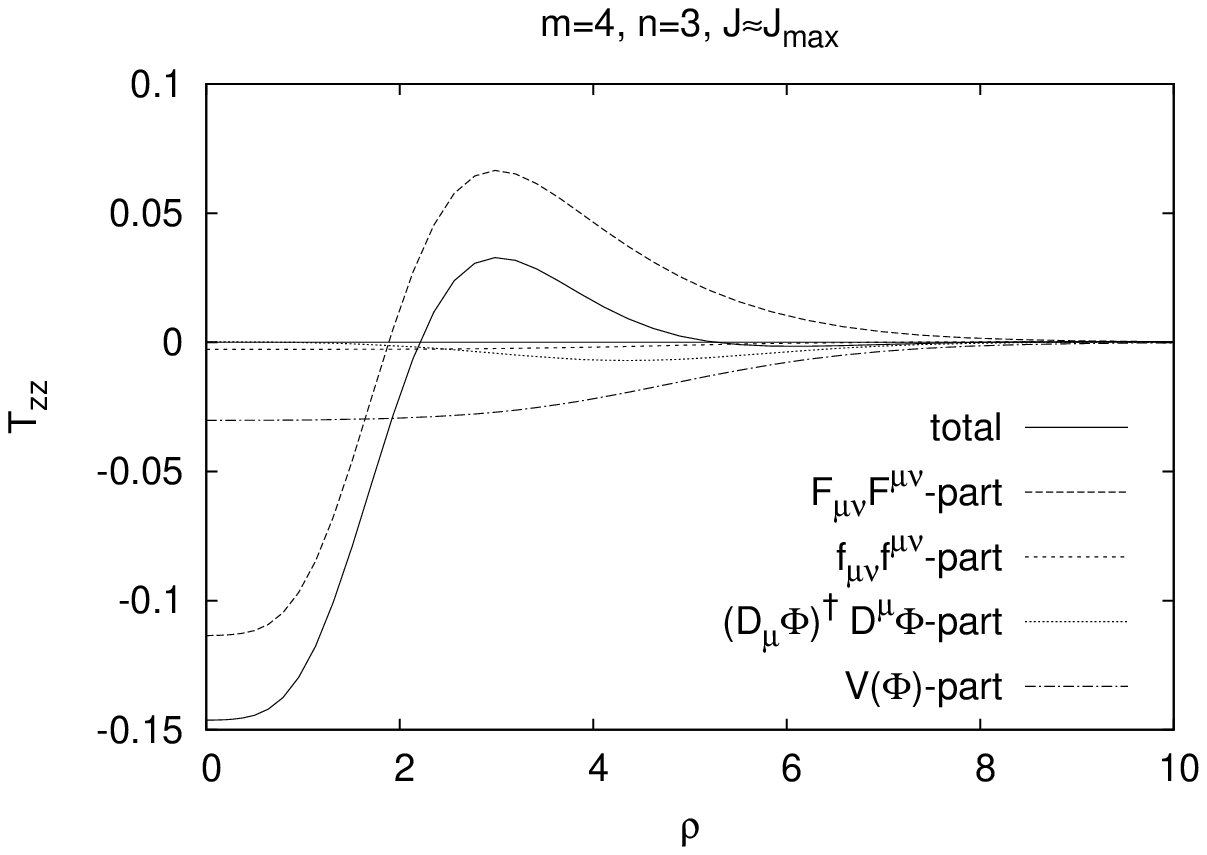}
\end{center}
\vspace{-0.5cm}
\caption{\small
The full stress-energy component $T_{zz}$ 
(upper)
and its $SU(2)$, $U(1)$, Higgs covariant derivative
and Higgs potential parts in
the equatorial plane (lower)
for the sphaleron-antisphaleron systems
$m=2$, $n=1$ $J=0$ (left)
and $m=4$, $n=3$, $J\approx J_{\rm max}$ (right).
}
\end{figure}

\section{Conclusions}

We have considered
sphaleron-antisphaleron pairs, chains and vortex ring solutions
in Weinberg-Salam theory,
which are characterized by two integers $n$ and $m$.
Starting from the respective neutral electroweak configurations,
we have obtained the corresponding 
branches of rotating electrically charged solutions.
These branches exist up to maximal values of the charge
and angular momentum,
beyond which localized solutions are no longer possible.

The angular momentum $J$ and the charge $\cal Q$ of these
sphaleron-antisphaleron systems are proportional
$$J = n {\cal Q} /e.$$
Their energy and binding energy increase with increasing rotation,
and so does their magnetic moment.
With increasing charge
the energy density of the configurations spreads further out,
while its overall magnitude reduces.
At the same time the effect of the rotation 
is a centrifugal shift of the energy density tori
to larger radii.

We have also addressed the equilibrium condition (\ref{Tzz})
for these sphaleron-antisphaleron systems. 
In all systems, it is the surface integral that vanishes
to give equilibrium, and not the 
stress-energy tensor component $T_{zz}$ by itself.
However, for the sphaleron-antisphaleron pair
(and other even $m$ chains)
the stress-energy tensor component $T_{zz}$
almost vanishes in the equatorial plane.
In these configurations the positive contribution from the
$SU(2)$ part almost cancels the negative
contributions from the $U(1)$ and Higgs parts,
thus yielding an almost vanishing total $T_{zz}$
in the equatorial plane.

The next step will be to study the fermion modes
in the background of such rotating electroweak configurations
to understand their relevance for baryon number violating
processes \cite{Kunz:1993ir}.
Moreover it will be interesting to include the effect of gravitation
\cite{Ibadov:2008hj} to obtain rotating gravitating regular configurations
as well as black hole solutions.

{\bf Acknowledgement}:
We gratefully acknowledge discussions with G.~Gibbons,
E.~Radu and M.~S.~Volkov.
R.I.~acknowledges support by the Volkswagen Foundation,
and B.K.~support by the DFG.



\begin{thebibliography}{000}

\bibitem{Manton:1983nd}
  N.~S.~Manton,
  Phys.\ Rev.\  D {\bf 28}, 2019 (1983).

\bibitem{Klinkhamer:1984di}
  F.~R.~Klinkhamer and N.~S.~Manton,
  Phys.\ Rev.\  D {\bf 30}, 2212 (1984).

\bibitem{Kleihaus:1991ks}
  B.~Kleihaus, J.~Kunz and Y.~Brihaye,
  Phys.\ Lett.\  B {\bf 273}, 100 (1991).

\bibitem{Kunz:1992uh}
  J.~Kunz, B.~Kleihaus and Y.~Brihaye,
  Phys.\ Rev.\  D {\bf 46}, 3587 (1992).

\bibitem{'tHooft:1976up}
  G.~'t Hooft,
  Phys.\ Rev.\ Lett.\  {\bf 37}, 8 (1976).

\bibitem{Ringwald:1989ee}
  A.~Ringwald,
  Nucl.\ Phys.\  B {\bf 330}, 1 (1990).

\bibitem{McLerran:1993rv}
  L.~D.~McLerran,
  Acta Phys.\ Polon.\  B {\bf 25}, 309 (1994)
  [arXiv:hep-ph/9311239].

\bibitem{Rubakov:1996vz}
  V.~A.~Rubakov and M.~E.~Shaposhnikov,
   ``Electroweak baryon number non-conservation in the early universe and in
  Usp.\ Fiz.\ Nauk {\bf 166}, 493 (1996)
  [Phys.\ Usp.\  {\bf 39}, 461 (1996)]
  [arXiv:hep-ph/9603208].

\bibitem{Dine:2003ax}
  M.~Dine and A.~Kusenko,
  Rev.\ Mod.\ Phys.\  {\bf 76}, 1 (2004)
  [arXiv:hep-ph/0303065].

\bibitem{Klinkhamer:2003hz}
  F.~R.~Klinkhamer and C.~Rupp,
  J.\ Math.\ Phys.\  {\bf 44}, 3619 (2003)
  [arXiv:hep-th/0304167].

\bibitem{Saffin:1997ae}
  P.~M.~Saffin and E.~J.~Copeland,
  Phys.\ Rev.\  D {\bf 57}, 5064 (1998)
  [arXiv:hep-ph/9710343].

\bibitem{Radu:2008ta}
  E.~Radu and M.~S.~Volkov,
  Phys.\ Rev.\  D {\bf 79}, 065021 (2009)
  [arXiv:0810.0908 [hep-th]].

\bibitem{Kleihaus:2008cv}
  B.~Kleihaus, J.~Kunz and M.~Leissner,
  Phys.\ Lett.\  B {\bf 678}, 313 (2009)
  [arXiv:0810.1142 [hep-ph]].

\bibitem{Kleihaus:1994yj}
  B.~Kleihaus and J.~Kunz,
  Phys.\ Lett.\  B {\bf 329}, 61 (1994)
  [arXiv:hep-ph/9403289].

\bibitem{Kleihaus:1994tr}
  B.~Kleihaus and J.~Kunz,
  Phys.\ Rev.\  D {\bf 50}, 5343 (1994)
  [arXiv:hep-ph/9405387].

\bibitem{Brihaye:1994ib}
  Y.~Brihaye and J.~Kunz,
  Phys.\ Rev.\  D {\bf 50}, 4175 (1994)
  [arXiv:hep-ph/9403392].

\bibitem{Klinkhamer:1985ki}
  F.~R.~Klinkhamer,
  Z.\ Phys.\  C {\bf 29}, 153 (1985).

\bibitem{Klinkhamer:1990ik}
  F.~R.~Klinkhamer,
  Phys.\ Lett.\  B {\bf 246}, 131 (1990).

\bibitem{Klinkhamer:1993hb}
  F.~R.~Klinkhamer,
  Nucl.\ Phys.\  B {\bf 410}, 343 (1993)
  [arXiv:hep-ph/9306295].

\bibitem{Kleihaus:2008gn}
  B.~Kleihaus, J.~Kunz and M.~Leissner,
  Phys.\ Lett.\  B {\bf 663}, 438 (2008)
  [arXiv:0802.3275 [hep-th]].

\bibitem{Kleihaus:2003nj}
  B.~Kleihaus, J.~Kunz and Y.~Shnir,
  Phys.\ Lett.\  B {\bf 570}, 237 (2003)
  [arXiv:hep-th/0307110].

\bibitem{Kleihaus:2003xz}
  B.~Kleihaus, J.~Kunz and Y.~Shnir,
  Phys.\ Rev.\  D {\bf 68}, 101701 (2003)
  [arXiv:hep-th/0307215].

\bibitem{Kleihaus:2004is}
  B.~Kleihaus, J.~Kunz and Y.~Shnir,
  Phys.\ Rev.\  D {\bf 70}, 065010 (2004)
  [arXiv:hep-th/0405169].

\bibitem{VanderBij:2001nm}
  J.~J.~Van der Bij and E.~Radu,
  Int.\ J.\ Mod.\ Phys.\  A {\bf 17}, 1477 (2002)
  [arXiv:gr-qc/0111046].

\bibitem{Kleihaus:2002tc}
  B.~Kleihaus, J.~Kunz and F.~Navarro-Lerida,
  Phys.\ Rev.\ Lett.\  {\bf 90}, 171101 (2003)
  [arXiv:hep-th/0210197].

\bibitem{Volkov:2003ew}
  M.~S.~Volkov and E.~Wohnert,
  Phys.\ Rev.\  D {\bf 67}, 105006 (2003)
  [arXiv:hep-th/0302032].

\bibitem{Radu:2008pp}
  E.~Radu and M.~S.~Volkov,
  Phys.\ Rept.\  {\bf 468}, 101 (2008)
  [arXiv:0804.1357 [hep-th]].


\bibitem{Kleihaus:2007vf}
  B.~Kleihaus, J.~Kunz, F.~Navarro-Lerida and U.~Neemann,
  Gen.\ Rel.\ Grav.\  {\bf 40}, 1279 (2008)
  [arXiv:0705.1511 [gr-qc]].

\bibitem{Kleihaus:2003tn}
  B.~Kleihaus, J.~Kunz and K.~Myklevoll,
  Phys.\ Lett.\  B {\bf 582}, 187 (2004)
  [arXiv:hep-th/0310300].


\bibitem{FIDISOL}
 W. Sch\"onauer and R. Wei\ss ,
 J. Comput. Appl. Math. 27, 279 (1989).



\bibitem{Aharonov:1992wf}
  Y.~Aharonov, A.~Casher, S.~R.~Coleman and S.~Nussinov,
  Phys.\ Rev.\  D {\bf 46}, 1877 (1992).

\bibitem{Beig:2008qi}
  R.~Beig and R.~M.~Schoen,
  Class.\ Quant.\ Grav.\  {\bf 26}, 075014 (2009)
  [arXiv:0811.1727 [gr-qc]].

\bibitem{Beig:2009jd}
  R.~Beig, G.~W.~Gibbons and R.~M.~Schoen,
  Class.\ Quant.\ Grav.\  {\bf 26}, 225013 (2009)
  [arXiv:0907.1193 [gr-qc]].

\bibitem{Kunz:1993ir}
  J.~Kunz and Y.~Brihaye,
  Phys.\ Lett.\  B {\bf 304}, 141 (1993)
  [arXiv:hep-ph/9302313].

\bibitem{Ibadov:2008hj}
  R.~Ibadov, B.~Kleihaus, J.~Kunz and M.~Leissner,
  Phys.\ Lett.\  B {\bf 663}, 136 (2008)
  [arXiv:0802.3335 [gr-qc]].



\end{thebibliography}
\end{document}